\documentclass[usenatbib]{mn2e}
\usepackage{amsmath,fleqn}
\usepackage{epsf}

\newcommand\bcdot{{\bmath\cdot}}
\newcommand\btimes{{\bmath\times}}
\newcommand\bcolon{{\bmath:}}
\newcommand\grad{{\bmath\nabla}}

\newcommand\bme{{\bmath e}}
\newcommand\bmH{{\bmath H}}
\newcommand\bmn{{\bmath n}}
\newcommand\bmx{{\bmath x}}
\newcommand\bmS{{\bmath S}}
\newcommand\bmX{{\bmath X}}
\renewcommand\bmu{{\bmath u}}
\newcommand\bmv{{\bmath v}}
\newcommand\bmw{{\bmath w}}
\newcommand\rmd{\mathrm{d}}
\newcommand\rme{\mathrm{e}}
\newcommand\rmi{\mathrm{i}}
\newcommand\rmD{\mathrm{D}}
\newcommand\f{\frac}
\newcommand\p{\partial}
\newcommand\cst{\mathrm{constant}}

\title[An affine model of the dynamics of astrophysical discs]
{An affine model of the dynamics of astrophysical discs}

\author[Gordon I.\ Ogilvie]
{Gordon I.\ Ogilvie\\
Department of Applied Mathematics and Theoretical Physics,
University of Cambridge, Centre for Mathematical Sciences,\\
Wilberforce Road, Cambridge CB3 0WA}

\begin{document}

\maketitle

\label{firstpage}
 
\begin{abstract}
  Thin astrophysical discs are very often modelled using the equations
  of two-dimensional hydrodynamics. We derive an extension of this
  model that describes more accurately the behaviour of a thin disc in
  the absence of self-gravity, magnetic fields and complex internal
  motions. The ideal fluid theory is derived directly from Hamilton's
  Principle for a three-dimensional fluid after making a specific
  approximation to the deformation gradient tensor. We express the
  equations in Eulerian form after projection on to a reference
  plane. The disc is thought of as a set of fluid columns, each of
  which is capable of a time-dependent affine transformation,
  consisting of a translation together with a linear transformation in
  three dimensions. Therefore, in addition to the usual
  two-dimensional hydrodynamics in the reference plane, the theory
  allows for a deformation of the midplane (as occurs in warped discs)
  and for the internal shearing motions that accompany such
  deformations. It also allows for the vertical expansions driven in
  non-circular discs by a variation of the vertical gravitational
  field around the horizontal streamlines, or by a divergence of the
  horizontal velocity. The equations of the affine model embody
  conservation laws for energy and potential vorticity, even for
  non-planar discs. We verify that they reproduce exactly the linear
  theories of three-dimensional warped and eccentric discs in a
  secular approximation. However, the affine model does not rely on
  any secular or small-amplitude assumptions and should be useful in
  more general circumstances.
\end{abstract}


\section{Introduction}
\label{s:intro}

Astrophysical discs, consisting of continuous matter in orbital motion
around a massive body, are found throughout the Universe on a variety
of lengthscales.  They are usually thin, having a small aspect ratio
$H/r\ll1$, where $H$ is a measure of the extent of the disc in the
`vertical' direction perpendicular to the orbital plane at radius~$r$.
The dynamics of thin discs is very often studied using two-dimensional
(2D) equations that neglect the vertical extent and vertical motion of
the disc.  However, this approximation is not generally valid, even in
the limit $H/r\ll1$.

Studies of wave propagation in astrophysical discs
\citep{1998ApJ...504..983L}, and of the dynamics of eccentric or
tidally distorted discs
\citep{2001MNRAS.325..231O,2002MNRAS.330..937O}, have shown that
problems that have traditionally been studied using 2D models have
quite different solutions when the internal vertical structure and
vertical motion of the disc are taken into account, even when $H/r$ is
small.  For example, \citet{2008MNRAS.388.1372O} found that the
prograde precession of elliptical discs observed around Be stars has a
natural explanation only when these effects are included.

The essential physics involved here is that the vertical structure of
a thin disc is not generally hydrostatic except in the simplest
situation of a steady, circular disc around a single mass.  When the
axial symmetry of this situation is broken by a free eccentricity of
the disc or by the presence of an orbital companion, vertical
oscillations are driven by the variation of the vertical gravitational
force around the horizontal streamlines and by the divergence of the
horizontal motion.

A complementary situation occurs if the reflectional symmetry of the
disc about the midplane is broken. In particular, if the disc is
warped such that the local orbital plane varies with $r$, horizontal
oscillatory flows are driven within the disc. These are especially
strong in Keplerian discs, leading to the remarkable properties of
warp propagation in these systems
\citep{1983MNRAS.202.1181P,1995ApJ...438..841P,1999MNRAS.304..557O}.

While fully three-dimensional (3D) treatments of these situations are
possible, and indeed necessary for the resolution of
magnetohydrodynamic or other turbulence within the disc, global 3D
simulations of realistically thin discs over the timescales of
interest remain extremely demanding.  The purpose of this paper is to
present and analyse a new model that augments the equations of 2D
hydrodynamics to include the additional degrees of freedom that are
needed for a description of discs that lack axial or reflectional
symmetry.  It takes into account the variable and non-hydrostatic
thickness of astrophysical discs and allows a general displacement of
the midplane of the disc from a reference plane.  Although the method
by which the equations are derived does not definitively establish
their domain of applicability, the model does have a satisfying
mathematical structure and internal consistency.  In particular, we
show that it implies conservation laws for energy and potential
vorticity that generalize those of 2D hydrodynamics. Furthermore, we
show that the equations correctly reproduce the linear hydrodynamics
of eccentric and warped 3D discs in the secular approximation.

Relevant previous work was carried out by \citet{1999MNRAS.304..674S},
who supplemented the 2D hydrodynamic equations with dynamical
equations for the vertical velocity and scaleheight. Their equations
are similar to those we derive below for the symmetric case
(Section~\ref{s:symmetric}) but do not conserve energy or potential
vorticity because the horizontal equation of motion was left in its 2D
form. Earlier, \citet{1988ApJ...331..838P} wrote a dynamical equation
for the thickness of an incompressible disc or ring, and
\citet{1981ApJ...245..274L} studied in detail the resonances and wave
emission associated with vertical compression of a disc in a binary
system.

The remainder of this paper is structured as follows. Starting from a
classical Lagrangian description of an ideal fluid
(Section~\ref{s:lagrangian}) in 3D, we restrict the degrees of freedom
of the fluid to those of a continuum of extended fluid columns
(Section~\ref{s:columnar}). We introduce a thin-disc approximation
(Section~\ref{s:thin}) and describe the vertical structure of the disc
(Section~\ref{s:vertical}). We then derive the equations of motion
from the approximated Lagrangian (Section~\ref{s:motion}) and write
them in Eulerian form by projecting them on to a reference plane
(Section~\ref{s:projected}). We discuss the conservation laws
(Section~\ref{s:conservation}) and rotational symmetry
(Section~\ref{s:rotation}) of the model. The equations and their
interpretation are considerably simplified in the case of a
reflectionally symmetric disc (Section~\ref{s:symmetric}). We then
derive the equilibrium conditions and linear perturbation theory for
an axisymmetric disc (Section~\ref{s:axisymmetric}). We apply this to
warped and eccentric discs (Sections
\ref{s:warped}--\ref{s:eccentric}) to show that the model exactly
reproduces the previously known 3D secular theories in the appropriate
limits. Our conclusions follow in Section~\ref{s:conclusions}.

\section{Lagrangian approach}
\label{s:lagrangian}

In this paper we consider the case of an ideal fluid, which is
inviscid and undergoes adiabatic thermodynamics.  The equation of
motion can then be derived from Hamilton's Principle using a
Lagrangian analysis of the motion \citep[e.g.][]{1988AnRFM..20..225S}.

We label the fluid elements according to their position vectors
$\bmx_0=(x_0,y_0,z_0)$ in an arbitrary reference state.  The reference
state could be an initial condition or an equilibrium configuration,
but this is not necessary. The quantities $(x_0,y_0,z_0)$ are material
or Lagrangian coordinates.

Let $\bmx(\bmx_0,t)=(x,y,z)$ be the position vector of a fluid element
in the dynamical state at time $t$.  The fluid velocity is
\begin{equation}
  \bmu=\f{\rmD\bmx}{\rmD t},
\end{equation}
where
\begin{equation}
  \f{\rmD}{\rmD t}=\left(\f{\p}{\p t}\right)_{\bmx_0}
\end{equation}
is the Lagrangian time-derivative. Let
\begin{equation}
  J_{ij}=\f{\p x_i}{\p x_{0j}}
\end{equation}
be the Jacobian matrix of the time-dependent map from the reference
state to the dynamical state, and let
\begin{equation}
  J_3=\det(J_{ij})=\left|\f{\p(\bmx)}{\p({\bmx_0})}\right|
\end{equation}
be the Jacobian determinant of this 3D map.  The quantity $J_{ij}$ is
known in continuum mechanics as the deformation gradient tensor.

A mass element of the fluid may be written as
\begin{equation}
  \rmd m=\rho\,\rmd^3\bmx=\rho_0\,\rmd^3\bmx_0,
\end{equation}
where $\rho(\bmx,t)$ is the mass density in the dynamical state and
$\rho_0(\bmx_0)$ is the mass density in the reference state.  Mass
conservation implies
\begin{equation}
  \rho=J_3^{-1}\rho_0
\end{equation}
and we require $J_3$ to be strictly positive.

The exact Lagrangian for a non-self-gravitating ideal fluid is
\begin{equation}
  L=\int\left(\f{1}{2}|\bmu|^2-\Phi-e\right)\rmd m,
\label{exact_lagrangian}
\end{equation}
where $\Phi(\bmx,t)$ is the (external) gravitational potential and
$e(v,s)$ is the specific internal energy, which depends on the
specific volume
\begin{equation}
  v=\f{1}{\rho}=J_3v_0
\end{equation}
and the specific entropy
\begin{equation}
  s=s_0,
\end{equation}
$v_0(\bmx_0)$ and $s_0(\bmx_0)$ being the specific volume and entropy
in the reference state.  Fluid elements preserve their specific
entropy in an ideal fluid flow.  The differential of $e(v,s)$
satisfies the fundamental thermodynamic identity
\begin{equation}
  \rmd e=T\,\rmd s-p\,\rmd v,
\end{equation}
where $T$ is the temperature and $p$ is the pressure. In particular,
for a perfect gas of constant adiabatic index $\gamma$, we have
\begin{equation}
  \rho=J_3^{-1}\rho_0,\qquad
  p=J_3^{-\gamma}p_0
\end{equation}
and
\begin{equation}
  e=\f{p}{(\gamma-1)\rho}=J_3^{-(\gamma-1)}e_0.
\end{equation}

Hamilton's Principle states that the action functional $S[\bmx]=\int
L\,\rmd t$ is stationary, leading to the Euler--Lagrange
equation\footnote{This is the standard Euler--Lagrange equation for
  several functions ($\bmx$) of several variables ($\bmx_0,t$).  In
  our notation, $\rmD/\rmD t$ represents the derivative with respect
  to $t$ when $\bmx_0$ is held constant.  Also $u_i$ corresponds to
  $\rmD x_i/\rmD t$ and $J_{ij}$ to $\p x_i/\p x_{0j}$.}
\begin{equation}
  \f{\p\mathcal{L}_3}{\p x_i}-\f{\rmD}{\rmD t}\f{\p\mathcal{L}_3}{\p u_i}-\f{\p}{\p x_{0j}}\f{\p\mathcal{L}_3}{\p J_{ij}}=0,
\end{equation}
where $L=\int\mathcal{L}_3\,\rmd^3\bmx_0$.  After division by
$-\rho_0$, this gives the desired equation of motion
\begin{equation}
  \f{\rmD u_i}{\rmD t}=-\f{\p\Phi}{\p x_i}-\f{1}{\rho}\f{\p p}{\p x_i}.
\label{du}
\end{equation}

The pressure term in this equation deserves some comment. In the
Lagrangian approach this term emerges initially in the form
\begin{equation}
  -\f{1}{\rho_0}\f{\p}{\p x_{0j}}(J_3^{-\gamma}p_0C_{ij}),
\end{equation}
where
\begin{equation}
  C_{ij}=\f{1}{2}\epsilon_{ikm}\epsilon_{jln}J_{kl}J_{mn}=\f{\p J_3}{\p J_{ij}}=J_3\f{\p x_{0j}}{\p x_i}
\end{equation}
is the cofactor of the element $J_{ij}$ of the Jacobian matrix. Using
the identity $\p C_{ij}/\p x_{0j}=0$ to extract the cofactor from the
bracket, and then the chain rule to convert derivatives with respect
to the Lagrangian variable $\bmx_0$ to those with respect to the
Eulerian variable $\bmx$, we obtain the form
\begin{equation}
  -\f{1}{\rho}\f{\p p}{\p x_i},
\end{equation}
as given above.

\section{Columnar elements and affine transformation}
\label{s:columnar}

Our aim is to reduce the dynamics of a thin (but generally non-planar)
3D disc to a 2D description by applying certain assumptions and
approximations.  By doing this at the level of the Lagrangian
function, we can ensure that the resulting theory is self-consistent
and embodies the appropriate conservation laws.  Our derivation is
similar in spirit (although very different in detail) to the
derivation of the shallow-water model of geophysical fluid dynamics by
\citet{1985JFM...157..519M}.

\begin{figure*}
\centerline{\epsfxsize16cm\epsfbox{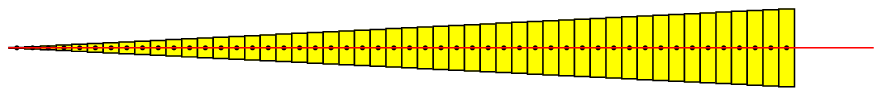}}
\centerline{\epsfxsize16cm\epsfbox{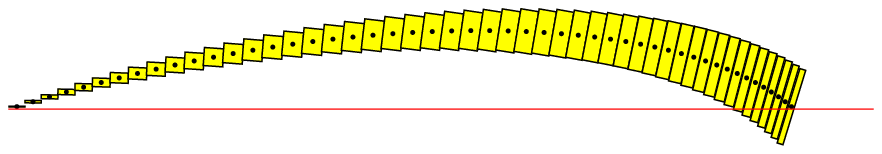}}
\caption{Illustration of the affine model, viewed in a cut
  perpendicular to the reference plane (red line). The disc is thought
  of as a continuum of extended fluid columns. In the reference state
  (top) the columns are vertical and their centres lie in the
  reference plane. To reach the dynamical state (bottom) the columns
  undergo time-dependent translations and linear transformations.}
\label{f:disc}
\end{figure*}

Although our disc is generally not planar, we will describe it
ultimately using a projection on to the plane $z=0$, which we regard
as horizontal and call the \textit{reference plane}.  In the case of a
central force, our model will have complete rotational symmetry and
the choice of reference plane is arbitrary.  We consider the disc to
be composed of extended fluid columns of infinitesimal width. The disc
is therefore regarded as a two-dimensional continuum of
one-dimensional elements (Fig.~\ref{f:disc}).

We envisage a convenient hypothetical reference state in which the
disc has axial and reflectional symmetry and is in vertical
hydrostatic equilibrium in a potential that has the same symmetries
(and which may differ from the actual potential $\Phi$). In the
reference state, the columnar elements are vertical and centred on the
reference plane $z=0$.  Let $H_0(r)$ be the vertical scaleheight (as
defined in Section~\ref{s:vertical} below) of the column whose centre
is at radius $r$ from the symmetry axis.

To reach the dynamical state of the disc, each fluid column may
undergo an arbitrary translation and an arbitrary linear
transformation in 3D.  The translation allows the centre of the column
to be moved to any point, not necessarily in the plane $z=0$.  The
linear transformation allows the column to be expanded or contracted
and also rotated about its centre.

The combination of a translation and a linear transformation is known
as an \textit{affine transformation}, which explains the name of our
model.  Note that each column undergoes an independent affine
transformation, the parameters of which will depend continuously on
the column label and also on time.

In the Lagrangian viewpoint, we label the columnar elements by the
horizontal position vectors $\bar\bmx_0=(x_0,y_0,0)$ of their centres
in the reference state. Generally, we use an overbar to denote a
planar quantity such as the horizontal projection of a 3D vector. The
3D fluid elements within each column are further identified by the
dimensionless label
\begin{equation}
  \zeta=\f{z_0}{H_0},
\end{equation}
which runs from $-\infty$ to $\infty$, with $\zeta=0$ corresponding to
the centre of the column and most of the mass being contained within
$|\zeta|<1$.  Under the affine transformation, the column maps to
\begin{equation}
  \bmx=\bmX(\bar\bmx_0,t)+\bmH(\bar\bmx_0,t)\zeta,
\label{XH}
\end{equation}
where $\bmX=(X,Y,Z)$ is the position vector of the centre of the
column in the dynamical state and $\bmH=(H_x,H_y,H_z)$ is a
\textit{scale vector} with the dimensions of length. For example, the
fluid elements labelled by $\zeta=\pm1$ are separated by
$2H_0\,\bme_z$ in the reference state but by $2\bmH$ in the dynamical
state.

The fluid velocity is then
\begin{equation}
  \bmu=\bmv+\bmw\zeta,
\end{equation}
where
\begin{equation}
  \bmv=\f{\rmD\bmX}{\rmD t}
\end{equation}
is the velocity of the centre of the column and
\begin{equation}
  \bmw=\f{\rmD\bmH}{\rmD t}
\end{equation}
is the rate of change of the scale vector. Note that $\rmD\zeta/\rmD
t=0$ because $\zeta$ is a Lagrangian coordinate labelling fluid
elements.

Each columnar element has six degrees of freedom $(\bmX,\bmH)$. The
variables $X$ and $Y$ give the fluid all the potentialities of
(compressible) 2D hydrodynamics. In addition, the variable $Z$ allows
the midplane of the disc to be deformed away from the plane $z=0$, as
occurs for example in warped discs. We refer to the surface $z=Z$,
which is the locus of column centres $\zeta=0$, as the
\textit{deformed midplane}.

The variable $H_z$ allows the disc to undergo vertical expansion or
contraction, as occurs for example in eccentric or tidally distorted
discs. Finally, the variables $H_x$ and $H_y$ allow the columns to be
tilted so that the disc undergoes internal shearing motions as in
warped discs.

It can be helpful to think of the map from the reference state to the
dynamical state as a composition of two stages:
$(x_0,y_0,z_0)\mapsto(X,Y,\zeta)\mapsto(x,y,z)$.  The Jacobian matrix
$J_{ij}$ and determinant $J_3$ of the composite map are the products
of those of the two stages.  The intermediate variables $(X,Y,\zeta)$
represent a system of `columnar' coordinates, with $\bar\bmX=(X,Y)$
identifying a column by means of the horizontal position vector of its
centre and $\zeta$ labelling the fluid elements within a column.  In
the Eulerian viewpoint we will regard quantities such as $Z$, $\bmH$,
$\bmv$ and $\bmw$ as functions of $(\bar\bmX,t)$ rather than functions
of $(\bar\bmx_0,t)$.

The first stage $(x_0,y_0,z_0)\mapsto(X,Y,\zeta)$ of the map has the
Jacobian matrix
\begin{equation}
  \f{\p(X,Y,\zeta)}{\p(x_0,y_0,z_0)}=\begin{pmatrix}\p X/\p x_0&\p X/\p y_0&0\\\p Y/\p x_0&\p Y/\p y_0&0\\-\zeta\p\ln H_0/\p x_0&-\zeta\p\ln H_0/\p y_0&1/H_0\end{pmatrix},
\label{j_first}
\end{equation}
with determinant
\begin{equation}
  \left|\f{\p(X,Y,\zeta)}{\p(x_0,y_0,z_0)}\right|=\f{J_2}{H_0},
\end{equation}
where
\begin{equation}
  J_2=\left|\f{\p(\bar\bmX)}{\p({\bar\bmx_0})}\right|=\left|\f{\p(X,Y)}{\p(x_0,y_0)}\right|
\end{equation}
is the Jacobian determinant of the 2D map $(x_0,y_0)\mapsto(X,Y)$ and
the factor of $1/H_0$ comes from $z_0\mapsto\zeta=z_0/H_0$.

In considering the second stage $(X,Y,\zeta)\mapsto(x,y,z)$ of the
map, it is helpful in preparation for an Eulerian viewpoint to regard
$Z$ and $\bmH$ as functions of $\bar\bmX$ rather than $\bar\bmx_0$, as
mentioned above.  The Jacobian matrix of the second stage is then
\begin{equation}
  \f{\p(x,y,z)}{\p(X,Y,\zeta)}=\begin{pmatrix}1+H_{x,X}\zeta&H_{x,Y}\zeta&H_x\\H_{y,X}\zeta&1+H_{y,Y}\zeta&H_y\\Z_X+H_{z,X}\zeta&Z_Y+H_{z,Y}\zeta&H_z\end{pmatrix},
\label{j_second}
\end{equation}
where $H_{x,X}=\p H_x/\p X$, $Z_X=\p Z/\p X$, etc.  Its determinant
will not generally be positive for all $\zeta$ because of the clashing
of neighbouring columns. In other words, the columnar coordinate
system generally breaks down sufficiently far from the (deformed)
midplane. However, for a thin disc with moderate deformations, we
expect there to be negligible mass in these distant regions. Indeed,
in the next Section we will make an approximation that prohibits the
Jacobian determinant from changing sign far from the disc.

The deformed midplane at any instant of time can now be thought of as
the surface $z=Z(X,Y)$ or $\bmx=\bmX(\bar\bmX)$. We assume that the
deformation of the disc is sufficiently moderate that $Z$ is a
single-valued function of $(X,Y)$, and that $J_2>0$. In some cases
this may require the reference plane to be chosen judiciously.  The
vector area element of the deformed midplane is
\begin{equation}
  \rmd\bmS=\left(\f{\p\bmX}{\p X}\btimes\f{\p\bmX}{\p Y}\right)\rmd X\,\rmd Y=\bmn\,\rmd X\,\rmd Y,
\label{dS}
\end{equation}
where the vector
\begin{equation}
  \bmn=\bme_z-\bar\grad Z=\left(-Z_X,-Z_Y,1\right)
\end{equation}
is normal to the surface, and generally of greater than unit length.
The notation
\begin{equation}
  \bar\grad=\left(\f{\p}{\p X},\f{\p}{\p Y},0\right)
\end{equation}
represents the planar gradient operator for quantities that depend on
$(\bar\bmX,t)$.

\section{Thin-disc approximation}
\label{s:thin}

\begin{figure}
\centerline{\epsfxsize8cm\epsfbox{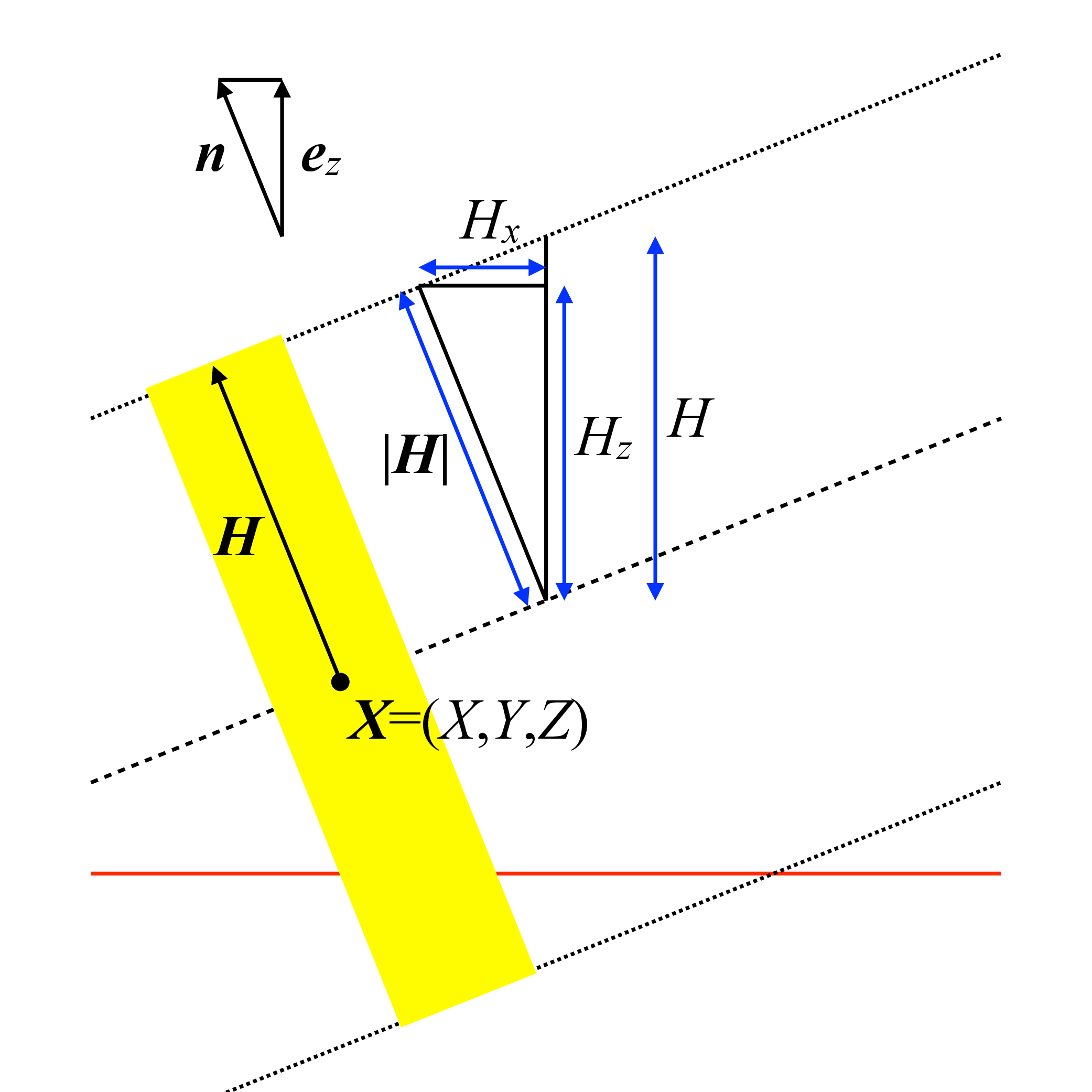}}
\vskip1cm
\caption{Illustration of the projected vertical scaleheight $H$ of the
  disc in the simple case of a flat disc that is tilted with respect
  to the reference plane (red solid line).  The dashed and dotted
  lines represent the deformed midplane $\zeta=0$ and the surfaces
  $\zeta=\pm1$ within which most of the mass is contained,
  respectively.  The yellow rectangle represents a single columnar
  element.}
\label{f:H}
\end{figure}

For a thin disc with large-scale deformations, we approximate the
Jacobian matrices (\ref{j_first}) and (\ref{j_second}) of the two
stages by evaluating them at $\zeta=0$ and neglecting their dependence
on $\zeta$.  Thus
\begin{equation}
  \f{\p(X,Y,\zeta)}{\p(x_0,y_0,z_0)}=\begin{pmatrix}\p X/\p x_0&\p X/\p y_0&0\\\p Y/\p x_0&\p Y/\p y_0&0\\0&0&1/H_0\end{pmatrix},
\end{equation}
with determinant
\begin{equation}
  \left|\f{\p(X,Y,\zeta)}{\p(x_0,y_0,z_0)}\right|=\f{J_2}{H_0}
\end{equation}
as before, and
\begin{equation}
  \f{\p(x,y,z)}{\p(X,Y,\zeta)}=\begin{pmatrix}1&0&H_x\\0&1&H_y\\Z_X&Z_Y&H_z\end{pmatrix},
\end{equation}
with determinant
\begin{equation}
  \left|\f{\p(x,y,z)}{\p(X,Y,\zeta)}\right|=H_z-\bar\bmH\bcdot\bar\grad Z=\bmH\bcdot\bmn.
\end{equation}
We define the important quantity
\begin{equation}
  H=\bmH\bcdot\bmn,
\end{equation}
which is the projected vertical scaleheight of the disc
(Fig.~\ref{f:H}).

This approximation results in a deformation gradient tensor that is
uniform within each column, and equal to the exact expression at the
centre of each column.  It can be justified on scaling grounds if
$\|\bar\grad\bmH\|\ll1$, i.e.\ if $|\bmH|$ is small compared to the
lengthscale on which $\bmH$ varies.  This condition should be
satisfied in a thin disc if the deformations are of large scale.

Under this approximation, the Jacobian determinant of the composite
map is
\begin{equation}
  J_3=J_2\f{H}{H_0}.
\label{j3j2}
\end{equation}

The $(3,3)$ element of the approximated inverse Jacobian matrix is
\begin{equation}
  \left(\f{\p z_0}{\p z}\right)_{x,y}=\f{J_2}{J_3}=\f{H_0}{H}.
\end{equation}
Therefore a vertical integration through the disc at constant $(x,y)$
becomes
\begin{equation}
  \int_{-\infty}^\infty\cdot\,\rmd z=\int_{-\infty}^\infty\cdot\,\f{H}{H_0}\,\rmd z_0=H\int_{-\infty}^\infty\cdot\,\rmd\zeta.
\label{vint}
\end{equation}
In other words, in order to remain at constant $x$ and $y$ as we
increase $\zeta$, we must sample different fluid columns if they are
tilted. Equation~(\ref{XH}) tells us that $X$ and $Y$ must change such
that $\rmd X=-H_x\,\rmd\zeta$ and $\rmd Y=-H_y\,\rmd\zeta$. Therefore
$\rmd z=\rmd Z+H_z\,\rmd\zeta=H\,\rmd\zeta$ with
$H=H_z-\bar\bmH\bcdot\bar\grad Z$.

\section{Vertical structure}
\label{s:vertical}

Let $\Sigma$ and $P$ denote the density and pressure integrated
vertically (i.e.\ with respect to the coordinate perpendicular to the
reference plane).  In the reference state, their values are
\begin{equation}
  \Sigma_0(\bar\bmx_0)=\int\rho_0\,\rmd z_0,\qquad
  P_0(\bar\bmx_0)=\int p_0\,\rmd z_0.
\label{reference}
\end{equation}
The hydrostatic reference state may be written as
\begin{equation}
  \rho_0=\f{\Sigma_0}{H_0}F_\rho(\zeta),
\end{equation}
\begin{equation}
  p_0=\f{P_0}{H_0}F_p(\zeta),
\end{equation}
where the dimensionless functions $F_\rho$ and $F_p$ satisfy the
dimensionless equations of vertical structure,
\begin{equation}
  \f{\rmd F_p}{\rmd\zeta}=-F_\rho\zeta,
\end{equation}
\begin{equation}
  \int_{-\infty}^\infty F_\rho\,\rmd\zeta=1,
\end{equation}
\begin{equation}
  \int_{-\infty}^\infty F_p\,\rmd\zeta=1,
\end{equation}
The first of these equations is a dimensionless form of hydrostatic
balance in any gravitational field that is proportional to the height
above the midplane, which is generic for a non-self-gravitating thin
disc.  The second and third equations are normalization conditions
required for equation~(\ref{reference}). The first and second
dimensionless moments of the density are
\begin{equation}
  \int_{-\infty}^\infty F_\rho\zeta\,\rmd\zeta=0,
\end{equation}
which follows from the reflectional symmetry about the midplane, and
\begin{equation}
  \int_{-\infty}^\infty F_\rho\zeta^2\,\rmd\zeta=1,
\end{equation}
which follows from the equations of vertical structure after an
integration by parts. In dimensional terms we have
\begin{equation}
  H_0^2=\int\rho_0 z_0^2\,\rmd z_0\bigg/\int\rho_0\,\rmd z_0,
\end{equation}
which gives a precise meaning to the scaleheight as the standard
deviation of the density distribution.

Simple examples of solutions of these equations
\citep{2014MNRAS.445.2621O} are the isothermal structure,
\begin{equation}
  F_\rho(\zeta)=F_p(\zeta)=(2\pi)^{-1/2}\exp\left(-\f{\zeta^2}{2}\right),
\end{equation}
the homogeneous structure,
\begin{equation}
  F_\rho(\zeta)=\f{1}{2\sqrt{3}},
\end{equation}
\begin{equation}
  F_p(\zeta)=\f{3-\zeta^2}{4\sqrt{3}}
\end{equation}
(for $\zeta^2<3$ only), and the polytropic structure,
\begin{equation}
  F_\rho(\zeta)=C_n\left(1-\f{\zeta^2}{2n+3}\right)^n,
\end{equation}
\begin{equation}
  F_p(\zeta)=\f{2n+3}{2(n+1)}C_n\left(1-\f{\zeta^2}{2n+3}\right)^{n+1}
\end{equation}
(for $\zeta^2<2n+3$ only), where $n>0$ (not necessarily an integer) is the
polytropic index and
\[
  C_n=[(2n+3)\pi]^{-1/2}\f{\Gamma(n+{\textstyle\f{3}{2}})}{\Gamma(n+1)}
\]
is a normalization constant.  It can be shown that the polytropic
structure approaches the isothermal structure in the limit
$n\to\infty$, and approaches the homogeneous structure in the limit
$n\to0$. The reason for the multiplicity of possible solutions is that
either the vertical temperature profile, or the vertical entropy
profile, can be freely chosen in the case of an ideal fluid. In a
dissipative disc these profiles would be determined from a balance
between heating and cooling in the thermal energy equation.

An important property of the affine transformation is that each
columnar element undergoes a uniform expansion or compression, because
(in the thin-disc approximation explained in Section~\ref{s:thin}) the
Jacobian determinant $J_3$ is independent of $\zeta$. Therefore the
dimensionless profile of density is preserved, and so are those of
pressure and other thermodynamic variables if (as we assume here) the
gas is perfect and behaves adiabatically.

The density, pressure and specific internal energy of a perfect gas in
the dynamical state are therefore
\begin{equation}
  \rho=J_3^{-1}\rho_0,\qquad
  p=J_3^{-\gamma}p_0,\qquad
  e=J_3^{-(\gamma-1)}e_0.
\end{equation}
It follows from equations (\ref{vint}) and (\ref{j3j2}) that the
vertically integrated density and pressure are \begin{equation}
  \Sigma=J_2^{-1}\Sigma_0,\qquad P=J_2^{-1}J_3^{-(\gamma-1)}P_0.
\end{equation}
We can then write
\begin{equation}
  \rho=\bar\rho F_\rho(\zeta),\qquad
  p=\bar p F_p(\zeta),
\end{equation}
where
\begin{equation}
  \bar\rho=\f{\Sigma}{H},\qquad
  \bar p=\f{P}{H}
\end{equation}
are the representative density and pressure of each column.  Since
$F_\rho(0)$ varies between $1/2\sqrt{3}\approx0.289$ and
$1/\sqrt{2\pi}\approx0.399$, while $F_p(0)$ varies between
$\sqrt{3}/4\approx0.433$ and $1/\sqrt{2\pi}\approx0.399$, depending on
the polytropic index, the representative density and pressure are
larger by a factor of about $2$ or $3$ than the density and pressure
on the deformed midplane $\zeta=0$.

The scaleheight in the dynamical state is defined by
\begin{equation}
  H^2=\int\rho(z-Z)^2\,\rmd z\bigg/\int\rho\,\rmd z
\end{equation}
(where the integrals are carried out at constant $x$ and $y$), so it
is again the standard deviation of the density distribution
perpendicular to the reference plane.

Let $\mathcal{R}$ be the gas constant and $\mu$ the mean molecular
weight.  Then the temperature is
\begin{equation}
  T=\f{\mu}{\mathcal{R}}\f{p}{\rho}=\bar TF_T(\zeta),
\end{equation}
where
\begin{equation}
  \bar T=\f{\mu}{\mathcal{R}}\f{\bar p}{\bar\rho}=\f{\mu}{\mathcal{R}}\f{P}{\Sigma}
\end{equation}
and
\begin{equation}
  F_T=\f{F_p}{F_\rho}.
\end{equation}
The specific entropy is (apart from an unimportant additive constant)
\begin{equation}
  s=\f{\mathcal{R}}{\mu}\f{1}{\gamma-1}\ln\left(p\rho^{-\gamma}\right)=\bar s+\f{\mathcal{R}}{\mu}\ln F_s(\zeta),
\label{entropy}
\end{equation}
where
\begin{equation}
  \bar s=\f{\mathcal{R}}{\mu}\f{1}{\gamma-1}\ln\left(\bar p\bar\rho^{-\gamma}\right)=\f{\mathcal{R}}{\mu}\f{1}{\gamma-1}\ln\left(P\Sigma^{-\gamma}H^{\gamma-1}\right)
\end{equation}
and
\begin{equation}
  F_s=F_p^{1/(\gamma-1)}F_\rho^{-\gamma/(\gamma-1)}.                                
\end{equation}

\section{Lagrangian and equations of motion}
\label{s:motion}

We now express the Lagrangian (\ref{exact_lagrangian}) of the ideal
fluid in terms of the variables we have introduced.

The 3D mass element is
\begin{equation}
  \rmd m=\rho_0\,\rmd^3\bmx_0=\rho_0\,\rmd^2\bar\bmx_0\,\rmd z_0=\Sigma_0\,\rmd^2\bar\bmx_0\,F_\rho\,\rmd\zeta.
\end{equation}

For the kinetic energy, we have
\begin{eqnarray}
  \lefteqn{\int\f{1}{2}|\bmu|^2\,\rmd m}&\nonumber\\
  &&=\iint\left(\f{1}{2}|\bmv|^2+\bmv\bcdot\bmw\,\zeta+\f{1}{2}|\bmw|^2\zeta^2\right)\Sigma_0\,\rmd^2\bar\bmx_0\,F_\rho\,\rmd\zeta\nonumber\\
  &&=\int\f{1}{2}\left(|\bmv|^2+|\bmw|^2\right)\Sigma_0\,\rmd^2\bar\bmx_0.
\end{eqnarray}

For the gravitational energy, we expand the gravitational potential in
a Taylor series about the centre of the fluid column:
\begin{eqnarray}
  \Phi(\bmx)&=&\Phi(\bmX+\bmH\zeta)\nonumber\\
  &=&\Phi(\bmX)+\zeta\bmH\bcdot\grad\Phi+\f{1}{2}\zeta^2\bmH\bmH\bcolon\grad\grad\Phi+\cdots
\end{eqnarray}
(in which the derivatives are evaluated at $\bmX$, and we have
suppressed any explicit time-dependence of the potential). For a thin
disc, we accept the (quadrupolar) truncation
\begin{equation}
  \int\Phi\,\rmd m=\int\left[\Phi(\bmX)+\f{1}{2}\bmH\bmH\bcolon\grad\grad\Phi\right]\Sigma_0\,\rmd^2\bar\bmx_0.
\end{equation}

Finally, for the internal energy of a perfect gas, we have
\begin{eqnarray}
  \int e\,\rmd m&=&\iint J_3^{-(\gamma-1)}e_0\,\Sigma_0\,\rmd^2\bar\bmx_0\,F_\rho\,\rmd\zeta\nonumber\\
  &=&\int J_3^{-(\gamma-1)}\f{p_0}{(\gamma-1)\rho_0}\,\Sigma_0\,\rmd^2\bar\bmx_0\,F_\rho\,\rmd\zeta\nonumber\\
  &=&\iint\f{J_3^{-(\gamma-1)}P_0}{(\gamma-1)}\,\rmd^2\bar\bmx_0\,F_p\,\rmd\zeta\nonumber\\
  &=&\int\f{J_3^{-(\gamma-1)}P_0}{(\gamma-1)}\,\rmd^2\bar\bmx_0.
\end{eqnarray}

Thus we obtain the Lagrangian
\begin{eqnarray}
  \lefteqn{L=\int\left[\f{1}{2}\left(|\bmv|^2+|\bmw|^2\right)-\Phi(\bmX,t)-\f{1}{2}\bmH\bmH\bcolon\grad\grad\Phi\right.}&\nonumber\\
  &&\left.\qquad-\f{J_3^{-(\gamma-1)}P_0}{(\gamma-1)\Sigma_0}\right]\Sigma_0\,\rmd^2\bar\bmx_0.
\end{eqnarray}
Writing this as $L=\int\mathcal{L}_2\,\rmd^2\bar\bmx_0$, where the
Lagrangian density $\mathcal{L}_2$ depends on $\bmX$ and $\bmH$ and
their derivatives with respect to $t$ and $\bar\bmx_0$, we identify
the Euler--Lagrange equations as
\begin{equation}
  \f{\p\mathcal{L}_2}{\p X_i}-\f{\rmD}{\rmD t}\f{\p\mathcal{L}_2}{\p v_i}-\f{\p}{\p\bar x_{0j}}\f{\p\mathcal{L}_2}{\p(\p X_i/\p\bar x_{0j})}=0,
\end{equation}
\begin{equation}
  \f{\p\mathcal{L}_2}{\p H_i}-\f{\rmD}{\rmD t}\f{\p\mathcal{L}_2}{\p w_i}-\f{\p}{\p\bar x_{0j}}\f{\p\mathcal{L}_2}{\p(\p H_i/\p\bar x_{0j})}=0,
\end{equation}
where summation over $j=\{1,2\}$ is implied.  After division by
$-\Sigma_0$ and application of algebraic identities, these give the
desired equations of motion
\begin{equation}
  \f{\rmD^2\bmX}{\rmD t^2}=-\grad\Phi-\f{1}{2}\bmH\bmH\bcolon\grad\grad\grad\Phi-\f{1}{\Sigma}\bar\grad P+\f{1}{\Sigma}\bar\grad\bcdot\left(\f{P\bar\bmH\bmn}{H}\right),
\label{d2r}
\end{equation}
\begin{equation}
  \f{\rmD^2\bmH}{\rmD t^2}=-\bmH\bcdot\grad\grad\Phi+\f{P\bmn}{\Sigma H}.
\label{d2a}
\end{equation}
The terms in equation~(\ref{d2r}) involving the vertically integrated
pressure $P$ are written here in terms of derivatives with respect to
the Eulerian coordinates $(X,Y)$ on the reference plane, rather than
the Lagrangian coordinates $(x_0,y_0)$; this involves operations
similar to those leading to equation~(\ref{du}). In the last term, the
divergence is taken on the first index (belonging to $\bar\bmH$).  The
terms involving $\bmn$ in these equations, which are not present in 2D
hydrodynamics, come from the property that $J_3$ is proportional to
$H=\bmH\bcdot\bmn$.

\section{Projected Eulerian representation}
\label{s:projected}

We now interpret equations (\ref{d2r}) and (\ref{d2a}) fully in an
Eulerian sense, projected on to the reference plane $z=0$.  The
projected Eulerian form of the equations is
\begin{equation}
  \f{\rmD\bmv}{\rmD t}=-\grad\Phi-\f{1}{2}\bmH\bmH\bcolon\grad\grad\grad\Phi-\f{1}{\Sigma}\bar\grad P+\f{1}{\Sigma}\bar\grad\bcdot\left(\f{P\bar\bmH\bmn}{H}\right),
\label{dv}
\end{equation}
\begin{equation}
  \f{\rmD\bmw}{\rmD t}=-\bmH\bcdot\grad\grad\Phi+\f{P\bmn}{\Sigma H},
\label{dw}
\end{equation}
with
\begin{equation}
  \f{\rmD}{\rmD t}=\f{\p}{\p t}+\bar\bmv\bcdot\bar\grad,
\end{equation}
\begin{equation}
  \bmv=\bar\bmv+v_z\,\bme_z,
\end{equation}
\begin{equation}
  v_z=\f{\rmD Z}{\rmD t},
\end{equation}
\begin{equation}
  \bmw=\f{\rmD\bmH}{\rmD t},
\end{equation}
and again $H=\bmH\bcdot\bmn$ with $\bmn=\bme_z-\bar\grad Z$. In
addition we need evolutionary equations for $\Sigma$ and $P$. From
$\Sigma=J_2^{-1}\Sigma_0$ we obtain, as in 2D hydrodynamics,
\begin{equation}
  \f{\rmD\ln\Sigma}{\rmD t}=-\bar\grad\bcdot\bar\bmv.
\end{equation}
From $P=J_2^{-1}J_3^{-(\gamma-1)}P_0$ we find
\begin{equation}
  \f{\rmD\ln P}{\rmD t}=-\gamma\bar\grad\bcdot\bar\bmv-(\gamma-1)\f{\rmD\ln H}{\rmD t},
\end{equation}
in which
\begin{equation}
  \f{\rmD\ln H}{\rmD t}=\f{1}{H}\f{\rmD}{\rmD t}(\bmH\bcdot\bmn)=\f{1}{H}\bmn\bcdot(\bmw-\bar\bmH\bcdot\bar\grad\bmv).
\end{equation}

These equations have numerous alternative forms such as
\begin{equation}
  \f{\p\Sigma}{\p t}+\bar\grad\bcdot(\Sigma\bar\bmv)=0,
\end{equation}
\begin{equation}
  \f{\rmD\bar\rho}{\rmD t}=-\bar\rho\left(\bar\grad\bcdot\bar\bmv+\f{\rmD\ln H}{\rmD t}\right),
\end{equation}
\begin{equation}
  \f{\rmD\bar p}{\rmD t}=-\gamma\bar p\left(\bar\grad\bcdot\bar\bmv+\f{\rmD\ln H}{\rmD t}\right),
\end{equation}
\begin{equation}
  \f{\rmD\bar s}{\rmD t}=0,
\end{equation}
etc.

A full set of equations is written out explicitly in Cartesian
coordinates in Appendix~\ref{s:appendix2}. Polar coordinates would of
course be more appropriate for many applications.

An Eulerian representation of the fluid variables, valid within a few
scaleheights of the deformed midplane, is
\begin{equation}
  \bmu\approx\bmv(\bar\bmx,t)+\bmw(\bar\bmx,t)\zeta,
\end{equation}
\begin{equation}
  \rho\approx\bar\rho(\bar\bmx,t)F_\rho(\zeta),
\end{equation}
\begin{equation}
  p\approx\bar p(\bar\bmx,t)F_p(\zeta),
\end{equation}
where
\begin{equation}
  \zeta=\f{z-Z}{H},\qquad
  \bar\rho=\f{\Sigma}{H},\qquad
  \bar p=\f{P}{H}.
\end{equation}

Some care is needed with the notation of derivatives. In the terms of
equation (\ref{dv}) involving $P$, the operator $\bar\grad$ acts on
planar quantities that are functions of $(X,Y,t)$ only, and there is
no ambiguity concerning these derivatives. In contrast, $\Phi$ is
generally a function of $(x,y,z,t)$; the horizontal components of
$\grad\Phi$ in equation (\ref{dv}) are obtained by differentiating
$\Phi$ with respect to $x$ or $y$ and then setting $\bmx=\bmX$, rather
than by first evaluating the potential at $z=Z(X,Y,t)$ and then
differentiating with respect to $X$ or $Y$, which would introduce
further terms via the chain rule.

While equation~(\ref{dv}) contains all the terms present in 2D
hydrodynamics, it differs from that model in several
respects. Firstly, the equation has a vertical component, describing
how the midplane of the disc moves vertically in situations lacking
reflectional symmetry (e.g.\ a warped disc). Secondly, the second term
on the right-hand side is the gravitational quadrupolar force acting
on the extended fluid column; as seen in Section~\ref{s:axisymmetric}
below, this term is active even in a hydrostatic situation. Thirdly,
the last term on the right-hand side is a novel force arising from
pressure and a deformation of the midplane; this term conserves
momentum but leads to an anisotropic stress in the reference plane. It
may seem puzzling that an anisotropic stress can arise from
pressure. For example, the vertical component of equation~(\ref{dv})
indicates that there is a horizontal flux density of vertical momentum
equal to $-P\bar\bmH/H$ within the reference plane. In fact, the flux
density of vertical momentum in 3D is just $p\,\bme_z$; however, if
the columns are tilted then the pressure transmits vertical momentum
from one column to its neighbours, resulting in an apparent horizontal
flux within the reference plane.

Equation~(\ref{dw}) is relatively novel, although the vertical
component describes breathing oscillations of the disc and has been
considered in previous work \citep[e.g.][]{1999MNRAS.304..674S}. The
horizontal components capture the shearing horizontal oscillations
driven by pressure gradients in warped discs or other situations
lacking reflectional symmetry.

The thin-disc approximation introduced in Section~\ref{s:thin} was
justified on the grounds that $\|\bar\grad\bmH\|\ll1$, i.e.\ that
$|\bmH|$ is small compared to the lengthscale on which $\bmH$ varies.
This approximation results in a Lagrangian that does not depend on the
spatial derivatives of $\bmH$ and gives rise to the equations in the
form presented above.  In Appendix~\ref{s:appendix2} we present the
form of the equations for a more general model in which the Jacobian
determinant $J_3$ is allowed to depend on the spatial derivatives of
$\bmH$.  We will see in Section~\ref{s:axisymmetric} below that
particular extensions of this type are desirable to improve the
accuracy and stability of the model at small scales comparable to
$|\bmH|$.

\section{Conservation of energy and potential vorticity}
\label{s:conservation}

The equations of the previous Section imply the local conservation of
total energy in the Eulerian form
\begin{eqnarray}
  \lefteqn{\f{\p}{\p t}(\Sigma\mathcal{E})+\bar\grad\bcdot\left[(\Sigma\mathcal{E}+P)\bar\bmv-\f{P}{H}(\bmv\bcdot\bmn)\bar\bmH\right]}&\nonumber\\
  &&=\Sigma\left(\dot\Phi+\f{1}{2}\bmH\bmH\bcolon\grad\grad\dot\Phi\right),
\label{energy}
\end{eqnarray}
with specific total energy
\begin{equation}
  \mathcal{E}=\f{1}{2}\left(|\bmv|^2+|\bmw|^2\right)+\Phi+\f{1}{2}\bmH\bmH\bcolon\grad\grad\Phi+\f{P}{(\gamma-1)\Sigma}.
\end{equation}
This expression for $\mathcal{E}$ has a clear interpretation: the
first two terms are kinetic energy, the next two are gravitational
potential energy (again in the quadrupolar approximation for extended
fluid columns) and the last term is internal energy. The source term
on the right-hand side of equation~(\ref{energy}) involves
$\dot\Phi=\p\Phi/\p t$, which vanishes in the case of a
time-independent potential.

The fact that there is an exact form of energy conservation in the
affine model is reassuring and implies a certain self-consistency. It
is not surprising, however, because we derived the model from
Hamilton's Principle and the conservation of energy is directly
related to the symmetry of the Lagrangian under time translation.

Less obvious is the conservation of \textit{potential vorticity} (PV).
Also known as \textit{vortensity} in the context of astrophysical
discs, this is a modified version of the vertical component of
vorticity that is conserved in ideal, barotropic 2D hydrodynamics and
has been found to play an important role in numerous problems in
astrophysical discs.  In geophysical fluid dynamics, the theory of
potential vorticity is highly developed.  Expressions for the PV take
a variety of forms depending on the model (shallow-water,
quasi-geostrophic, etc.) being employed, but PV conservation can
always be related to Kelvin's circulation theorem and derived from the
symmetry of the Lagrangian under the continuous relabelling of fluid
elements \citep[e.g.][]{1985JFM...157..519M,2018BC}.

We define the PV in the affine model as
\begin{equation}
  q=\f{1}{\Sigma}\bmn\bcdot\left[\bar\grad\btimes(\bmv+w_i\bar\grad H_i)\right],
\end{equation}
where there is an implied summation over Cartesian indices
$i=\{1,2,3\}$.  Note that $\bmv$, and therefore
$\bar\grad\btimes\bmv$, are generally three-component vectors.  It can
then be shown from the equations of the preceding Section that
\begin{equation}
  \f{\rmD q}{\rmD t}=\f{S_q}{\Sigma}
\label{pv}
\end{equation}
or, in Eulerian conservative form,
\begin{equation}
  \f{\p}{\p t}(\Sigma q)+\bar\grad\bcdot(\Sigma q\bar\bmv)=S_q,
\end{equation}
where
\begin{equation}
  S_q=\left[\bar\grad\left(\f{P}{H}\right)\btimes\bar\grad\left(\f{H}{\Sigma}\right)\right]_z
\end{equation}
is a baroclinic source of PV per unit area.  The source term can be
written in various ways, e.g.
\begin{equation}
  S_q=[\bar\grad\bar p\btimes\bar\grad\bar v]_z=[\bar\grad\bar T\btimes\bar\grad\bar s]_z.
\end{equation}
Since the gradient vectors are horizontal and $n_z=1$, these
expressions are equivalent to
\begin{equation}
  S_q=\bmn\bcdot\left[\bar\grad\btimes(\bar p\bar\grad\bar v)\right]=\bmn\bcdot\left[\bar\grad\btimes(\bar T\bar\grad\bar s)\right].
\end{equation}

Consider a simple, closed material curve $C$ that lies in the deformed
midplane and moves with the velocity field $\bmv$.  Let $S$ be the
open material surface consisting of the region of the deformed
midplane enclosed by $C$.  The projections of $C$ and $S$ on the
reference plane are the planar curve $\bar C$ and the planar area
$\bar S$.  Integration of equation~(\ref{pv}) over $\bar S$ with
respect to the invariant mass element $\rmd m=\Sigma\,\rmd X\,\rmd Y$
results in
\begin{equation}
  \f{\rmd}{\rmd t}\int q\,\rmd m=\int_{\bar S}S_q\,\rmd X\,\rmd Y,
\end{equation}
i.e.
\begin{eqnarray}
  \lefteqn{\f{\rmd}{\rmd t}\int_{\bar S}\bmn\bcdot\left[\bar\grad\btimes(\bmv+w_i\bar\grad H_i)\right]\rmd X\,\rmd Y}&\nonumber\\
  &&=\int_{\bar S}\bmn\bcdot\left[\bar\grad\btimes(\bar T\bar\grad\bar s)\right]\rmd X\,\rmd Y.
\end{eqnarray}
Using expression~(\ref{dS}) for the vector area element, we may write this as
\begin{equation}
  \f{\rmd}{\rmd t}\int_S\left[\bar\grad\btimes(\bmv+w_i\bar\grad H_i)\right]\bcdot\,\rmd\bmS=\int_S\left[\bar\grad\btimes(\bar T\bar\grad\bar s)\right]\bcdot\,\rmd\bmS.
\end{equation}
By Stokes's theorem, this implies
\begin{equation}
  \f{\rmd}{\rmd t}\oint_C(\bmv+w_i\bar\grad H_i)\bcdot\rmd\bmX=\oint_C(\bar T\bar\grad\bar s)\bcdot\,\rmd\bmX.
\end{equation}
In particular, if $C$ is an isentropic material curve on which $\bar
s$ is constant, then we verify Kelvin's circulation theorem in the
form
\begin{equation}
  \f{\rmd}{\rmd t}\oint_C(\bmv+w_i\bar\grad H_i)\bcdot\,\rmd\bmX=0.
\end{equation}

The conserved circulation can also be written as
\begin{equation}
  \oint(\bmv\bcdot\,\rmd\bmX+\bmw\bcdot\,\rmd\bmH).
\end{equation}
This can be interpreted as the action integral
\begin{equation}
  \oint\sum_ip_i\,\rmd q_i
\end{equation}
of Hamiltonian dynamics, where $q_i$ are the generalized coordinates
(in our case, $\bmX$ and $\bmH$) and $p_i$ are the conjugate momenta
per unit mass (in our case, $\bmv$ and $\bmw$).  It can also be
related to the conserved circulation in 3D ideal hydrodynamics, which
is the line integral $\oint\bmu\bcdot\,\rmd\bmx$ around a closed
material curve within an isentropic surface.  Given the
expression~(\ref{entropy}) for the specific entropy in our disc of
non-zero thickness, if $\bar s$ is constant around~$C$ then the
isentropic material curves are those displaced from $C$ by any
constant value of $\zeta$.  On these curves, $\bmx=\bmX+\bmH\zeta$,
$\rmd\bmx=\rmd\bmX+\zeta\,\rmd\bmH$ and $\bmu=\bmv+\bmw\zeta$.
Expanding the differential
$\bmu\bcdot\,\rmd\bmx=(\bmv+\bmw\zeta)\bcdot(\rmd\bmX+\zeta\,\rmd\bmH)$
and replacing $\zeta$ and $\zeta^2$ with their mass-weighted averages
of $0$ and $1$, respectively, we plausibly obtain the above expression
$\bmv\bcdot\,\rmd\bmX+\bmw\bcdot\,\rmd\bmH$.

\section{Case of a central force}
\label{s:rotation}

For a central force deriving from a spherically symmetric potential
$\Phi(R)$, where $R=|\bmx|$, we have
\begin{equation}
  \grad\Phi=\f{1}{R}\f{\rmd\Phi}{\rmd R}\,\bmx,
\end{equation}
\begin{equation}
  \bmH\bcdot\grad\grad\Phi=\f{1}{R}\f{\rmd}{\rmd R}\left(\f{1}{R}\f{\rmd\Phi}{\rmd R}\right)(\bmH\bcdot\bmx)\bmx+\f{1}{R}\f{\rmd\Phi}{\rmd R}\,\bmH,
\end{equation}
\begin{eqnarray}
  \lefteqn{\bmH\bmH\bcolon\grad\grad\grad\Phi=\f{1}{R}\f{\rmd}{\rmd R}\left[\f{1}{R}\f{\rmd}{\rmd R}\left(\f{1}{R}\f{\rmd\Phi}{\rmd R}\right)\right](\bmH\bcdot\bmx)^2\bmx}&\nonumber\\
  &&+\f{1}{R}\f{\rmd}{\rmd R}\left(\f{1}{R}\f{\rmd\Phi}{\rmd R}\right)\left[2(\bmH\bcdot\bmx)\bmH+|\bmH|^2\bmx\right].
\end{eqnarray}
In particular, a Newtonian point-mass potential has
\begin{equation}
  \Phi=-\f{GM}{R},
\end{equation}
\begin{equation}
  \f{1}{R}\f{\rmd\Phi}{\rmd R}=\f{GM}{R^3},
\end{equation}
\begin{equation}
  \f{1}{R}\f{\rmd}{\rmd R}\left(\f{1}{R}\f{\rmd\Phi}{\rmd R}\right)=-\f{3GM}{R^5},
\end{equation}
\begin{equation}
  \f{1}{R}\f{\rmd}{\rmd R}\left[\f{1}{R}\f{\rmd}{\rmd R}\left(\f{1}{R}\f{\rmd\Phi}{\rmd R}\right)\right]=\f{15GM}{R^7}.
\end{equation}

Even though the equations of Section~\ref{s:projected} are projected
on to a reference plane, they do possess complete rotational symmetry
in the case of a central force, and would have the same form for any
choice of the reference plane.  We will verify this in
Section~\ref{s:warped} below through the demonstration of a rigid-tilt
mode of zero frequency.

\section{The symmetric case}
\label{s:symmetric}

An important special case occurs when the gravitational potential has
reflectional symmetry about the reference plane and the disc also
shares this symmetry.  It is helpful to introduce the notation
\begin{equation}
  \Psi=\f{\p^2\Phi}{\p z^2}\bigg|_{z=0}.
\end{equation}
Reflectional symmetry of the disc implies that $Z=0$, so $\bmn=\bme_z$
and $H=H_z$. The vectors $\bmH$ and $\bmw$ are purely vertical, while
$\bmv$ is purely horizontal. We can simplify the notation by writing
$w$ for $w_z$ and $\bmv$ for $\bar\bmv$.  The equations then reduce to
\begin{equation}
  \left(\f{\p}{\p t}+\bmv\bcdot\bar\grad\right)\bmv=-\bar\grad\Phi-\f{1}{2}H^2\bar\grad\Psi-\f{1}{\Sigma}\bar\grad P,
\label{dv_sym}
\end{equation}
\begin{equation}
  \left(\f{\p}{\p t}+\bmv\bcdot\bar\grad\right)w=-H\Psi+\f{P}{\Sigma H},
\label{dw_sym}
\end{equation}
\begin{equation}
  \left(\f{\p}{\p t}+\bmv\bcdot\bar\grad\right)H=w,
\end{equation}
together with appropriate equations for $\Sigma$ and $P$ (or
equivalent variables), e.g.
\begin{equation}
  \left(\f{\p}{\p t}+\bmv\bcdot\bar\grad\right)\Sigma=-\Sigma\bar\grad\bcdot\bmv,
\end{equation}
\begin{equation}
  \left(\f{\p}{\p t}+\bmv\bcdot\bar\grad\right)P=-\gamma P\bar\grad\bcdot\bmv-\f{(\gamma-1)Pw}{H}.
\end{equation}

We may write an explicit 3D Eulerian representation of the fluid
variables in this case as
\begin{equation}
  \bmu\approx\bmv(\bar\bmx,t)+w(\bar\bmx,t)\zeta\,\bme_z,
\end{equation}
\begin{equation}
  \rho\approx\bar\rho(\bar\bmx,t)F_\rho(\zeta),
\end{equation}
\begin{equation}
  p\approx\bar p(\bar\bmx,t)F_p(\zeta),
\end{equation}
where
\begin{equation}
  \zeta=\f{z}{H},\qquad
  \bar\rho=\f{\Sigma}{H},\qquad
  \bar p=\f{P}{H}.
\end{equation}

The specific energy and potential vorticity simplify to
\begin{equation}
  \mathcal{E}=\f{1}{2}\left(|\bmv|^2+w^2\right)+\Phi+\f{1}{2}H^2\Psi+\f{P}{(\gamma-1)\Sigma},
\end{equation}
\begin{equation}
  q=\f{1}{\Sigma}\bme_z\bcdot\left[\bar\grad\btimes(\bmv+w\bar\grad H)\right].
\end{equation}

\section{Axisymmetric equilibrium and linearized equations}
\label{s:axisymmetric}

If the potential is also steady and axisymmetric, such that $\Phi$ and
$\Psi$ are functions of cylindrical radius $r$ in the plane $z=0$,
then the simplest solution shares these symmetries, having
$\bmv=r\Omega(r)\,\bme_\phi$, $\Sigma=\Sigma(r)$, $P=P(r)$ and
$H=H(r)$, as well as $Z=0$, $\bmn=\bme_z$ and $\bmH=H\,\bme_z$ as in
the previous Section. Equations (\ref{dv_sym}) and (\ref{dw_sym}) give
\begin{equation}
  -r\Omega^2=-\f{\rmd\Phi}{\rmd r}-\f{1}{2}H^2\f{\rmd\Psi}{\rmd r}-\f{1}{\Sigma}\f{\rmd P}{\rmd r},
\end{equation}
\begin{equation}
  0=-H\Psi+\f{P}{\Sigma H}.
\end{equation}
The second of these equations corresponds to the vertical hydrostatic
equilibrium of the disc.  The first equation represents the radial
force balance, showing how the rotation of the disc differs from that
of a particle orbit because of the thickness and pressure of the disc.
The second term on the right-hand side of this equation does not
appear in 2D hydrodynamics, although it is generally comparable to the
third term; it represents the dilution of the radial gravitational
force due to the thickness of the disc, and can be interpreted as the
quadrupolar gravitational force acting on a fluid column.  Let
$\Omega_0(r)$ be the angular velocity of a circular particle orbit of
radius $r$, given by
\begin{equation}
  r\Omega_0^2=\f{\rmd\Phi}{\rmd r}\bigg|_{z=0}.
\end{equation}
The equilibrium conditions then reduce to
\begin{equation}
  P=\Sigma H^2\Psi,
\label{eqmz}
\end{equation}
\begin{equation}
  \f{\rmd}{\rmd r}(\Sigma H^2\Psi)+\f{1}{2}\Sigma H^2\f{\rmd\Psi}{\rmd r}=\Sigma r(\Omega^2-\Omega_0^2).
\label{eqmr}
\end{equation}

The linearized equations in the case of small departures from this
basic state separate into two decoupled subsystems.  The first is
relevant for perturbations that preserve the reflectional symmetry of
the disc, and takes the form
\begin{equation}
  \rmD v_r'-2\Omega v_\phi'=-HH'\f{\rmd\Psi}{\rmd r}-\f{1}{\Sigma}\f{\p P'}{\p r}+\f{\Sigma'}{\Sigma^2}\f{\rmd P}{\rmd r},
\end{equation}
\begin{equation}
  \rmD v_\phi'+\f{v_r'}{r}\f{\rmd}{\rmd r}(r^2\Omega)=-\f{1}{\Sigma r}\f{\p P'}{\p\phi},
\end{equation}
\begin{equation}
  \rmD\Sigma'+v_r'\f{\rmd\Sigma}{\rmd r}=-\f{\Sigma}{r}\left[\f{\p}{\p r}(rv_r')+\f{\p v_\phi'}{\p\phi}\right],
\end{equation}
\begin{equation}
  \rmD P'+v_r'\f{\rmd P}{\rmd r}=-\f{\gamma P}{r}\left[\f{\p}{\p r}(rv_r')+\f{\p v_\phi'}{\p\phi}\right]-\f{(\gamma-1)Pw'}{H},
\end{equation}
\begin{equation}
  \rmD w'=-H'\Psi+\left(\f{P}{\Sigma H}\right)',
\end{equation}
\begin{equation}
  \rmD H'+v_r'\f{\rmd H}{\rmd r}=w',
\end{equation}
where $(r,\phi)$ are polar coordinates on the reference plane and
\begin{equation}
  \rmD=\f{\p}{\p t}+\Omega\f{\p}{\p\phi}.
\end{equation}
Note that
\begin{equation}
  \left(\f{P}{\Sigma H}\right)'=\f{P}{\Sigma H}\left(\f{P'}{P}-\f{\Sigma'}{\Sigma}-\f{H'}{H}\right).
\end{equation}

The second subsystem describes perturbations that break the
reflectional symmetry of the disc, and takes the form
\begin{eqnarray}
  \lefteqn{\rmD^2Z'=-\Psi Z'-\f{1}{2}\Xi H^2Z'-\f{\rmd\Psi}{\rmd r}HH_r'}&\nonumber\\
  &&+\f{1}{\Sigma r}\left[\f{\p}{\p r}\left(\f{rPH_r'}{H}\right)+\f{\p}{\p\phi}\left(\f{PH_\phi'}{H}\right)\right],
\end{eqnarray}
\begin{eqnarray}
  \lefteqn{(\rmD^2-\Omega^2)H_r'-2\Omega\,\rmD H_\phi'=-\f{\rmd}{\rmd r}(r\Omega_0^2)H_r'-\f{\rmd\Psi}{\rmd r}HZ'}&\nonumber\\
  &&-\f{P}{\Sigma H}\f{\p Z'}{\p r},
\end{eqnarray}
\begin{equation}
  (\rmD^2-\Omega^2)H_\phi'+2\Omega\,\rmD H_r'=-\Omega_0^2H_\phi'-\f{P}{\Sigma Hr}\f{\p Z'}{\p\phi},
\end{equation}
where
\begin{equation}
  \Xi=\f{\p^4\Phi}{\p z^4}\bigg|_{z=0}.
\end{equation}

\begin{figure*}
\centerline{\epsfbox{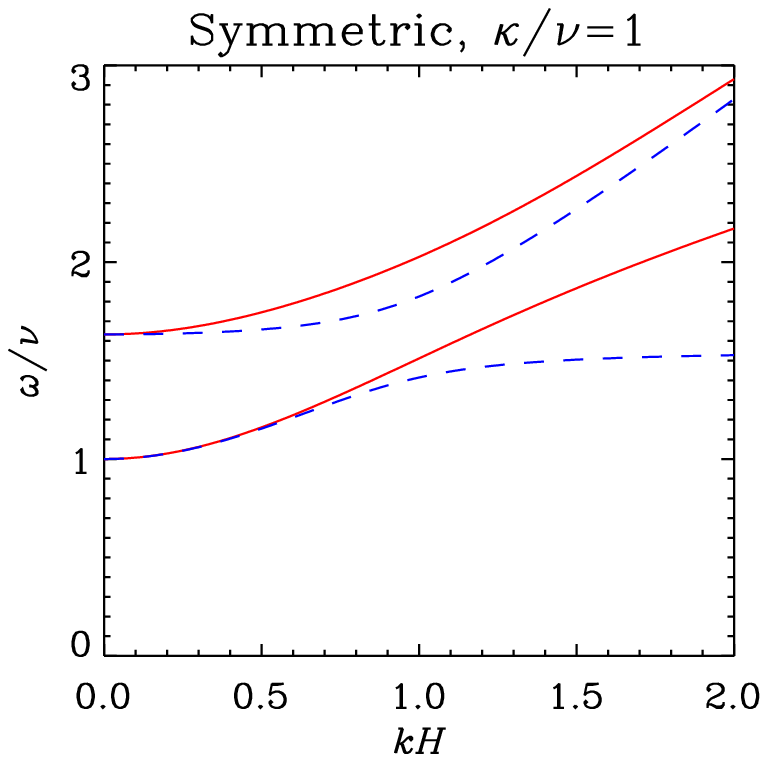}\epsfbox{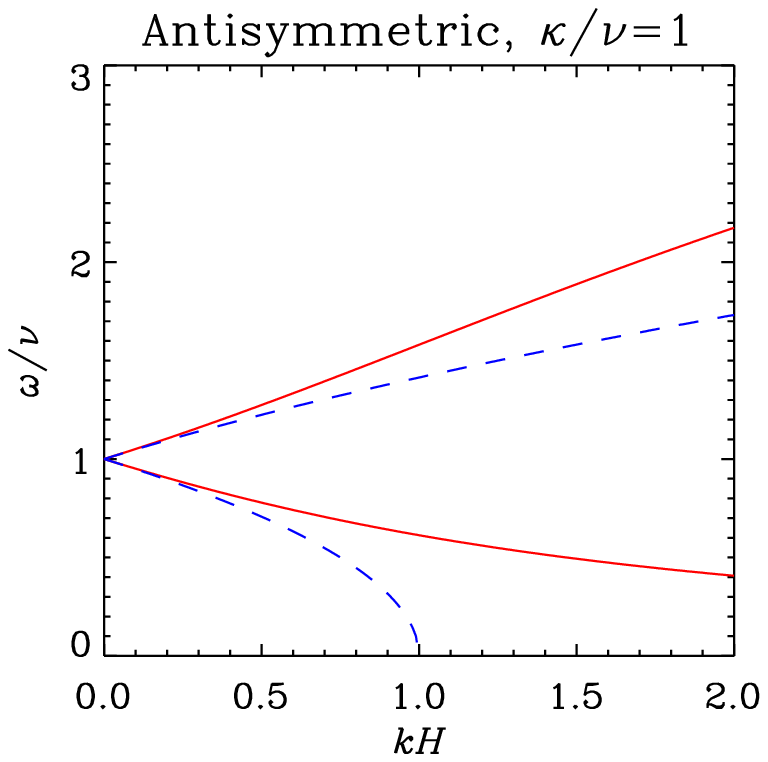}}
\centerline{\epsfbox{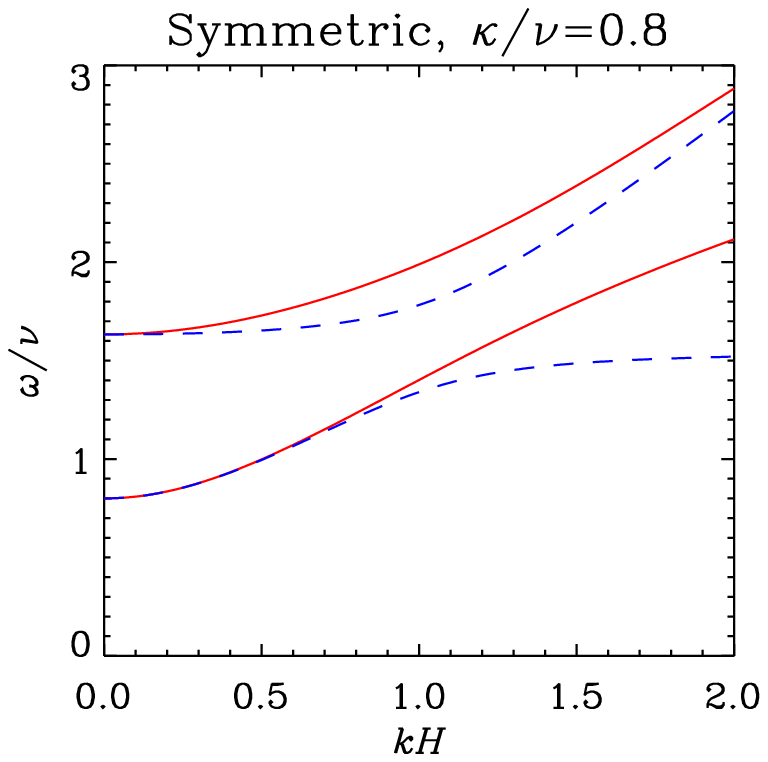}\epsfbox{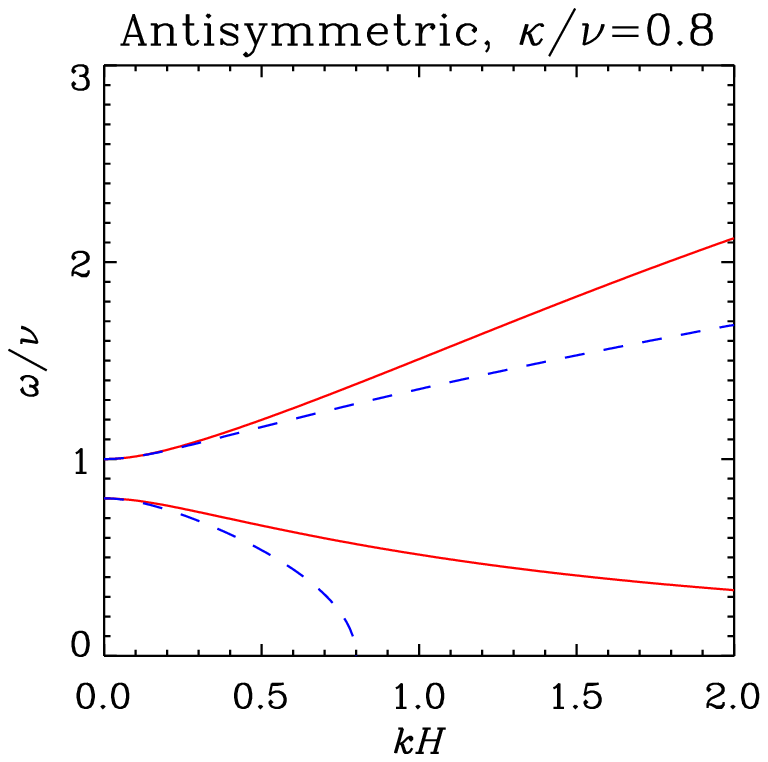}}
\caption{Local dispersion relation for a 3D polytropic disc (red solid
  lines) and in the unmodified affine model (blue dashed lines). In
  each case $\gamma=5/3$. The top panels are for a Keplerian disc and
  the lower two are for a non-Keplerian disc with $\kappa<\nu$. The
  left panels show the two symmetric modes and the right panels show
  the two antisymmetric modes. Other modes of the polytropic disc with
  higher vertical mode numbers are not plotted.}
\label{f:dispersion}
\end{figure*}

We will discuss special slowly varying solutions of the linearized
equations representing warped and eccentric discs in the following two
sections. A complementary situation is one in which the perturbations
have a short radial wavelength comparable to $H\ll r$. In this limit
the dominant variation of the perturbations is through the phase
factor
\begin{equation}
  \exp\left\{\rmi\left[\int k(r)\,\rmd r+m\phi-\omega t\right]\right\},
\end{equation}
where $k(r)$ is a local radial wavenumber satisfying $|k|r\gg 1$, $m$
(an integer of order unity) is an azimuthal wavenumber and $\omega$ is
an angular frequency. Let $\hat\omega=\omega-m\Omega$ be the intrinsic
wave frequency in the frame locally moving with the fluid. After some
algebra, we find that the local dispersion relation is
\begin{equation}
  \left(\hat\omega^2-\kappa^2-\f{\gamma P}{\Sigma}k^2\right)\left[\hat\omega^2-(\gamma+1)\nu^2\right]=\left[\f{(\gamma-1)Pk}{\Sigma H}\right]^2
\end{equation}
for symmetric modes and
\begin{equation}
  \left(\hat\omega^2-\kappa^2\right)\left(\hat\omega^2-\nu^2\right)=\left(\f{Pk}{\Sigma H}\right)^2
\end{equation}
for antisymmetric modes, where $\kappa$ and $\nu$ are the epicyclic
and vertical frequencies given by
\begin{equation}
  \kappa^2=\f{1}{r^3}\f{\rmd(r^4\Omega_0^2)}{\rmd r},
\end{equation}
\begin{equation}
  \nu^2=\Psi.
\end{equation}
Each case admits two solutions for $\hat\omega^2$. The symmetric case
involves a mixture of the classical density wave
$\hat\omega^2=\kappa^2+\f{\gamma P}{\Sigma}k^2$ with the breathing
mode $\hat\omega^2=(\gamma+1)\nu^2$; these are coupled when
$\gamma>1$. The antisymmetric case involves a coupling of the
epicyclic oscillation $\hat\omega^2=\kappa^2$ with the vertical
oscillation $\hat\omega^2=\nu^2$. Typical dispersion relations for the
case $\gamma=5/3$ are shown in Fig.~\ref{f:dispersion} where they are
compared with the corresponding modes in a 3D polytropic disc
(calculated as in \citealt{1995MNRAS.272..618K} or
\citealt{1998MNRAS.297..291O}). The polytropic disc is neutrally
stratified in order to eliminate internal gravity waves. It can be
seen from the figure that the affine model is accurate in describing
this type of motion for $kH\ll1$ and useful for $kH\la1$.

\begin{figure*}
\centerline{\epsfbox{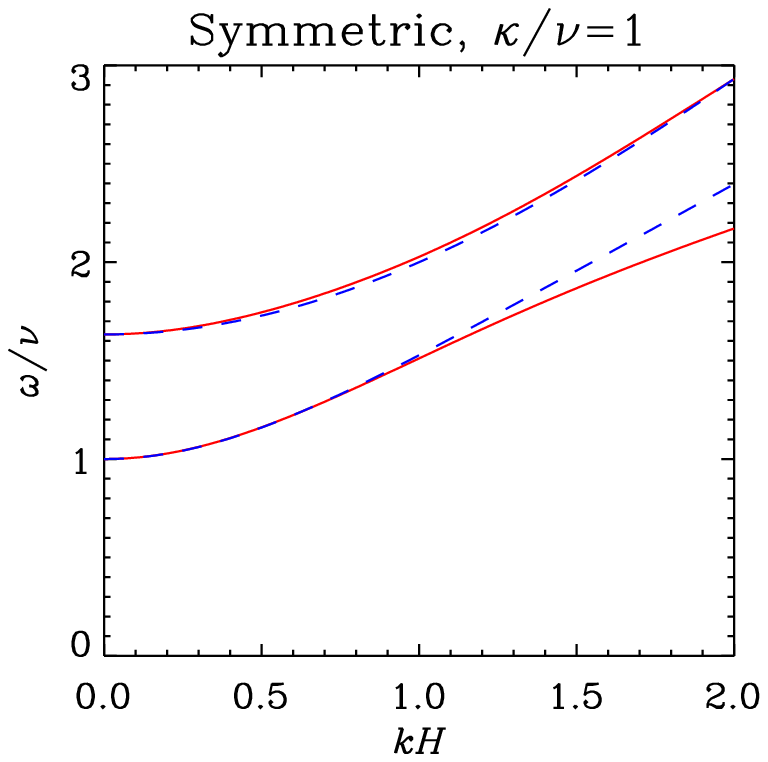}\epsfbox{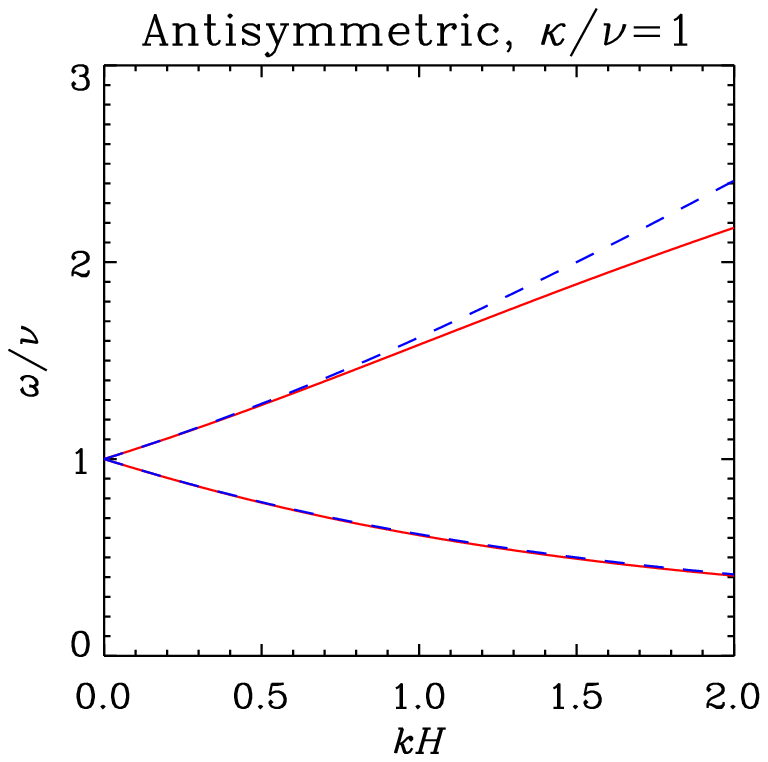}}
\centerline{\epsfbox{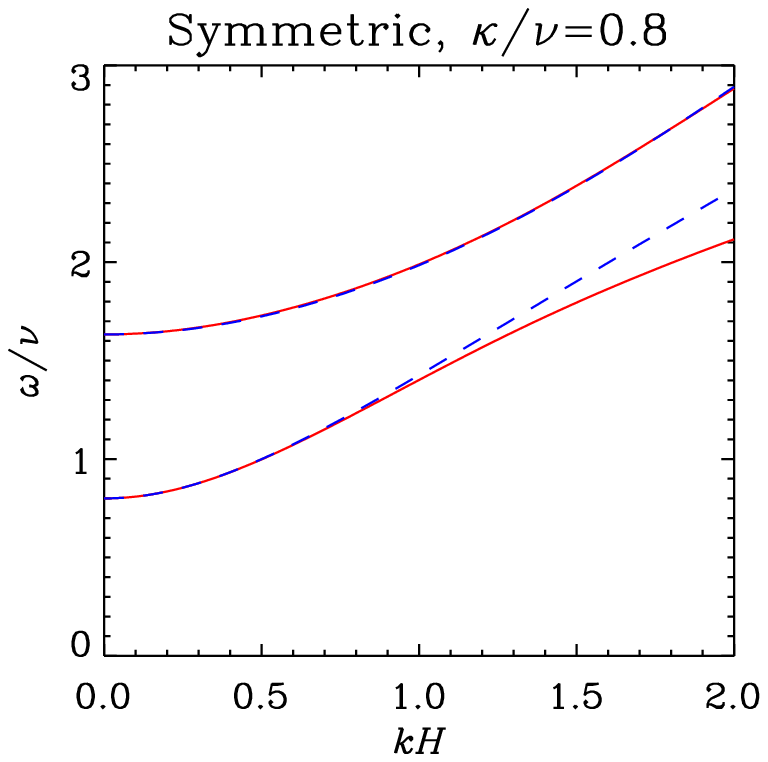}\epsfbox{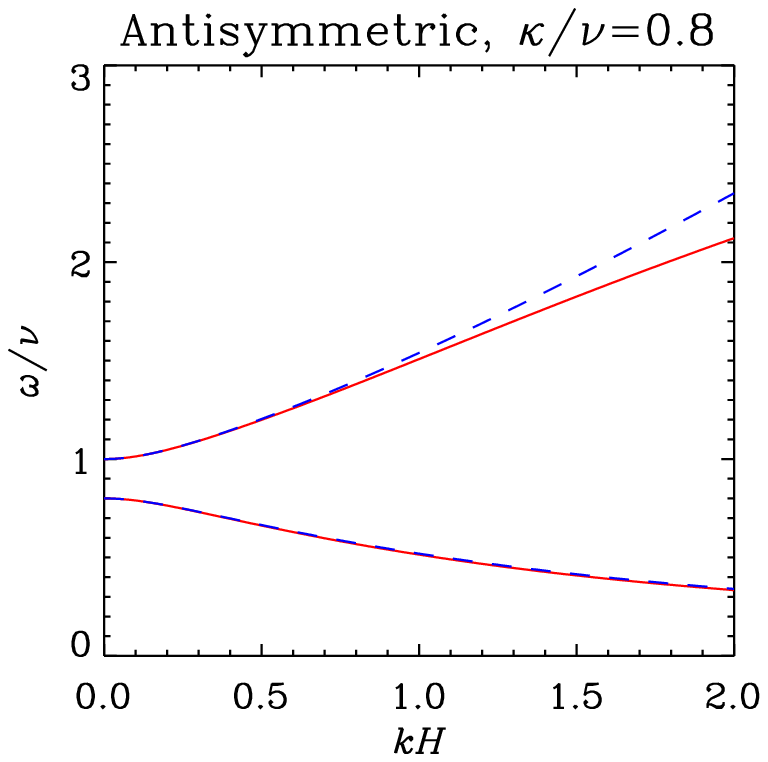}}
\caption{As for Fig.~\ref{f:dispersion} except that the affine model
  (blue dashed lines) has been modified as described in the text to
  improve the accuracy and stability of the model on small scales.}
\label{f:dispersion_modified}
\end{figure*}

It is hardly surprising that the dispersion relation is inaccurate for
$kH\gg1$.  In this limit the higher-frequency ($|\hat\omega|>\kappa$)
modes of a polytropic disc become concentrated near the surfaces of
the disc and the vertical structure of the velocity field is far
removed from the simple linear profile assumed in the affine model.
More concerning is the behaviour of the low-frequency antisymmetric
mode.  The smaller root for $\hat\omega^2$ vanishes at $kH=\kappa/\nu$
and becomes negative for larger $kH$, indicating instability on
wavelengths smaller than a few $H$.  This instability is unphysical
and needs to be suppressed in numerical implementations unless they
are of sufficiently low resolution.  Its origin can be traced to the
assumption made in Section~\ref{s:thin} that the deformation of the
disc is of large scale, leading to an approximation that makes the
internal energy insensitive to spatial derivatives of $\bmH$.  The
simple modification $F_1$ proposed in Appendix~\ref{s:appendix2}
restores such a dependence; when it is applied, we find that the
symmetric modes are unaffected, while the dispersion relation for
antisymmetric modes becomes
\begin{equation}
  \left(\hat\omega^2-\kappa^2-\f{P}{\Sigma}k^2\right)\left(\hat\omega^2-\nu^2\right)=\left(\f{Pk}{\Sigma H}\right)^2.
\end{equation}
This modification stabilizes the low-frequency antisymmetric mode at
large wavenumbers and in fact gives excellent agreement with the 3D
dispersion relation of the polytropic disc
(Fig.~\ref{f:dispersion_modified}); indeed it agrees exactly with the
dispersion relation of $n=1$ modes in a strictly isothermal disc.  It
is possible to improve the accuracy of the symmetric modes by making a
similar modification involving derivatives of $H_z$.  Including the
term $F_2$ proposed in Appendix~\ref{s:appendix2} changes the
dispersion relation for symmetric modes to
\begin{equation}
  \left(\hat\omega^2-\kappa^2-\f{\gamma P}{\Sigma}k^2\right)\left[\hat\omega^2-(\gamma+1)\nu^2-\f{P}{\Sigma}k^2\right]=\left[\f{(\gamma-1)Pk}{\Sigma H}\right]^2,
\end{equation}
which gives better agreement with the 3D dispersion relation of the
polytropic disc (Fig.~\ref{f:dispersion_modified}).

\section{Linear theory of warps}
\label{s:warped}

In this Section we assume that the potential is spherically symmetric,
which implies
\begin{equation}
  \Psi=\Omega_0^2,\qquad
  \Xi=\f{3}{r}\f{\rmd\Omega_0^2}{\rmd r}
\end{equation}
and eliminates nodal precession of inclined orbits.

For the antisymmetric perturbations, and assuming the azimuthal
dependence $\rme^{-\rmi\phi}$, we have
\begin{eqnarray}
  \lefteqn{\rmD^2Z'=-\Omega_0^2Z'-\f{\rmd\Omega_0^2}{\rmd r}\left(\f{3H^2Z'}{2r}+HH_r'\right)}\nonumber\\
&&+\f{1}{\Sigma r}\left[\f{\p}{\p r}\left(\f{rP H_r'}{H}\right)-\f{\rmi PH_\phi'}{H}\right],
\label{d2z'}
\end{eqnarray}
\begin{eqnarray}
  \lefteqn{(\rmD^2-\Omega^2)H_r'-2\Omega\,\rmD H_\phi'=-\f{\rmd(r\Omega_0^2)}{\rmd r}H_r'-\f{\rmd\Omega_0^2}{\rmd r}HZ'}\nonumber\\
  &&-\f{P}{\Sigma H}\f{\p Z'}{\p r},
\label{d2hr'}
\end{eqnarray}
\begin{equation}
  (\rmD^2-\Omega^2)H_\phi'+2\Omega\,\rmD H_r'=-\Omega_0^2H_\phi'+\f{\rmi P Z'}{\Sigma Hr},
\label{d2hp'}
\end{equation}
with now
\begin{equation}
  \rmD=\f{\p}{\p t}-\rmi\Omega.
\end{equation}
It is easily verified, using equations (\ref{eqmz}) and (\ref{eqmr}),
that these equations are exactly satisfied by a time-independent
rigid-tilt mode
\begin{equation}
  Z'=-r,\qquad
  H_r'=H,\qquad
  H_\phi'=-\rmi H,
\end{equation}
which corresponds to an infinitesimal change in the orientation of the
disc. This property is to be expected because of the complete
rotational symmetry of the problem.

Slowly varying warps in an inviscid disc have been treated by
\citet{1995ApJ...438..841P} and \citet{1999MNRAS.304..557O}, among
others. The behaviour is complicated by a resonance that occurs in
Keplerian discs owing to the coincidence of the orbital and epicyclic
frequencies. In the non-resonant case, the secular scalings for slowly
varying warps in a thin disc lead us to approximate $\Omega$ as
$\Omega_0$ and to neglect time-derivatives except where the leading
terms cancel in equation~(\ref{d2z'}):
\begin{eqnarray}
  \lefteqn{-2\rmi\Omega_0\f{\p Z'}{\p t}=\f{Z'}{\Sigma r\Omega_0}\f{\rmd}{\rmd r}(\Sigma H^2\Omega_0^3)-\f{\rmd\Omega_0^2}{\rmd r}\left(\f{3H^2Z'}{2r}+HH_r'\right)}\nonumber\\
&&+\f{1}{\Sigma r}\left[\f{\p}{\p r}\left(\f{rP H_r'}{H}\right)-\f{\rmi PH_\phi'}{H}\right],
\end{eqnarray}
\begin{equation}
  -\Omega_0^2H_r'+2\rmi\Omega_0^2H_\phi'=-\f{\rmd\Omega_0^2}{\rmd r}(rH_r'+HZ')-\Omega_0^2H\f{\p Z'}{\p r},
\end{equation}
\begin{equation}
  -\Omega_0^2H_\phi'-2\rmi\Omega_0^2H_r'=\f{\rmi\Omega_0^2HZ'}{r}.
\end{equation}
$H_\phi'$ can be eliminated to obtain
\begin{equation}
  -2\rmi\Omega_0\f{\p Z'}{\p t}=\f{\Omega_0^2Z'}{\Sigma}\f{\rmd}{\rmd r}\left(\f{\Sigma H^2}{r}\right)+\f{r\Omega_0^2}{\Sigma}\f{\p}{\p r}\left(\f{\Sigma HH_r'}{r}\right),
\end{equation}
\begin{equation}
  \f{\rmd(r^3\Omega_0^2)}{\rmd r}\f{H_r'}{H}=-\f{\p}{\p r}\left(r^3\Omega_0^2\f{Z'}{r}\right).
\end{equation}
These combine into
\begin{equation}
  2\rmi\Sigma r^3\Omega_0\f{\p W}{\p t}=\f{\p}{\p r}\left[\f{\Sigma H^2r^5\Omega_0^4}{\rmd(r^3\Omega_0^2)/\rmd r}\f{\p W}{\p r}\right],
\end{equation}
where $W=-Z'/r$ is the dimensionless tilt variable (related to the
inclination angle) used by \citet{1983MNRAS.202.1181P},
\citet{1995ApJ...438..841P} and others.  We see again that a
stationary rigid tilt ($W=\cst$) is a possible solution.  This
Schr\"odinger-like dispersive wave equation for the warp is exactly
equivalent to equation~(131) derived by \citet{1999MNRAS.304..557O}
from a global asymptotic analysis.

In the resonant case for a Keplerian disc ($\Omega_0\propto
r^{-3/2}$), equations (\ref{d2hr'}) and (\ref{d2hp'}) become
degenerate, both reducing to $H_\phi'\approx-2\rmi H_r'$ at leading
order.  Taking a $(1,2\rmi)$ linear combination of these equations to
eliminate the dominant terms, we obtain the approximation
\begin{equation}
  -2\rmi\Omega_0\f{\p H_r'}{\p t}=-r\Omega_0^2H\f{\p}{\p r}\left(\f{Z'}{r}\right)
\end{equation}
as well as
\begin{equation}
  -2\rmi\Omega_0\f{\p Z'}{\p t}=\f{r\Omega_0^2}{\Sigma}\f{\p}{\p r}\left(\f{\Sigma HH_r'}{r}\right).
\end{equation}
Identifying $Z'$ with $-rW$ (as above) and $\Sigma H\Omega_0^2r^2H_r'$
with $2\rmi G$, where $G$ is a complex internal torque variable, we
obtain exactly equations (5) and (6) of \citet{2002MNRAS.337..706L}
for an inviscid Keplerian disc, i.e.
\begin{equation}
  \Sigma r^2\Omega_0\f{\p W}{\p t}=\f{1}{r}\f{\p G}{\p r},
\end{equation}
\begin{equation}
  \f{\p G}{\p t}=\f{1}{4}\Sigma H^2r^3\Omega_0^3\f{\p W}{\p r},
\end{equation}
which combine into a non-dispersive wave equation for $W$, with wave
speed $H\Omega_0/2$.

\section{Linear theory of eccentric discs}
\label{s:eccentric}

We return to the linearized equations of Section~\ref{s:axisymmetric}
in the case of a point-mass potential for which
$\Omega_0^2=\Psi=GM/r^3$.  To make a comparison between the affine
model and the known secular theory of eccentric discs, we introduce
the small parameter $\epsilon\ll1$ such that $H/r=O(\epsilon)$, and
use it to expand the quantities of the basic state as
\begin{equation}
  \Omega=\Omega_0+\epsilon^2\Omega_2+\cdots,
\end{equation}
\begin{equation}
  \Sigma=\Sigma_0+\epsilon^2\Sigma_2+\cdots,
\end{equation}
\begin{equation}
  P=\epsilon^2(P_0+\epsilon^2P_2+\cdots),
\end{equation}
\begin{equation}
  H=\epsilon(H_0+\epsilon^2H_2+\cdots).
\end{equation}
The equilibrium conditions (\ref{eqmz}) and (\ref{eqmr}) reduce at
leading order to
\begin{equation}
  P_0=\Sigma_0H_0^2\Omega_0^2,
\end{equation}
\begin{equation}
  \f{\rmd P_0}{\rmd r}-\f{3P_0}{2r}=2\Sigma_0r\Omega_0\Omega_2.
\end{equation}

We describe a small eccentricity by considering reflectionally
symmetric perturbations proportional to $\rme^{-\rmi\phi}$.  The
linearized equations are
\begin{equation}
  \rmD v_r'-2\Omega v_\phi'=\f{3\Omega_0^2HH'}{r}-\f{1}{\Sigma}\f{\p P'}{\p r}+\f{\Sigma'}{\Sigma^2}\f{\rmd P}{\rmd r},
\end{equation}
\begin{equation}
  \rmD v_\phi'+\f{v_r'}{r}\f{\rmd(r^2\Omega)}{\rmd r}=\f{\rmi P'}{\Sigma r},
\end{equation}
\begin{equation}
  \rmD\Sigma'+v_r'\f{\rmd\Sigma}{\rmd r}=-\f{\Sigma}{r}\left[\f{\p}{\p r}(rv_r')-\rmi v_\phi'\right],
\end{equation}
\begin{equation}
  \rmD P'+v_r'\f{\rmd P}{\rmd r}=-\f{\gamma P}{r}\left[\f{\p}{\p r}(rv_r')-\rmi v_\phi'\right]-\f{(\gamma-1)Pw'}{H},
\end{equation}
\begin{equation}
  \rmD w'=-\Omega_0^2H'+\left(\f{P}{\Sigma H}\right)',
\end{equation}
\begin{equation}
  \rmD H'+v_r'\f{\rmd H}{\rmd r}=w',
\end{equation}
with
\begin{equation}
  \rmD=\f{\p}{\p t}-\rmi\Omega.
\end{equation}
We then expand
\begin{equation}
  v_r'=v_{r0}'+\epsilon^2v_{r2}'+\cdots,
\end{equation}
\begin{equation}
  v_\phi'=v_{\phi0}'+\epsilon^2v_{\phi2}'+\cdots,
\end{equation}
\begin{equation}
  w'=\epsilon w_0'+\cdots,
\end{equation}
\begin{equation}
  \Sigma'=\Sigma_0'+\cdots,
\end{equation}
\begin{equation}
  P'=\epsilon^2(P_0'+\cdots),
\end{equation}
\begin{equation}
  H'=\epsilon(H_0'+\cdots),
\end{equation}
where the perturbations depend on time through a slow variable
$\tau=\epsilon^2t$. The horizontal components of the equation of
motion at leading order are
\begin{equation}
  -\rmi\Omega_0v_{r0}'-2\Omega_0v_{\phi0}'=0,
\end{equation}
\begin{equation}
  -\rmi\Omega_0v_{\phi0}'+\f{1}{2}\Omega_0v_{r0}'=0,
\end{equation}
with solution
\begin{equation}
  v_{r0}'=\rmi r\Omega_0E(r,\tau),\qquad
  v_{\phi0}'=\f{1}{2}r\Omega_0E(r,\tau),
\end{equation}
representing a small eccentricity in the orbital motion.  Here $E$ is
the complex eccentricity used by \citet{2001MNRAS.325..231O} and
others. The remaining equations at leading order are
\begin{equation}
  -\rmi\Omega_0\Sigma_0'+\rmi r\Omega_0E\f{\rmd\Sigma_0}{\rmd r}=-\Sigma_0\,\rmi r\Omega_0\f{\p E}{\p r},
\end{equation}
\begin{equation}
  -\rmi\Omega_0P_0'+\rmi r\Omega_0E\f{\rmd P_0}{\rmd r}=-\gamma P_0\,\rmi r\Omega_0\f{\p E}{\p r}-\f{(\gamma-1)P_0w_0'}{H_0},
\end{equation}
\begin{equation}
  -\rmi\Omega_0w_0'=-\Omega_0^2H_0'+\f{P_0}{\Sigma_0H_0}\left(\f{P_0'}{P_0}-\f{\Sigma_0'}{\Sigma_0}-\f{H_0'}{H_0}\right),
\end{equation}
\begin{equation}
  -\rmi\Omega_0H_0'+\rmi r\Omega_0E\f{\rmd H_0}{\rmd r}=w_0',
\end{equation}
which have the solution
\begin{equation}
  \Sigma_0'=r\f{\p(\Sigma_0E)}{\p r},
\end{equation}
\begin{equation}
  P_0'=Er\f{\rmd P_0}{\rmd r}+\f{P_0}{\gamma}\left[3(\gamma-1)E+(2\gamma-1)r\f{\p E}{\p r}\right],
\end{equation}
\begin{equation}
  w_0'=\f{\rmi\Omega_0H_0}{\gamma}\left[3E-(\gamma-1)r\f{\p E}{\p r}\right],
\end{equation}
\begin{equation}
  H_0'=Er\f{\rmd H_0}{\rmd r}-\f{H_0}{\gamma}\left[3E-(\gamma-1)r\f{\p E}{\p r}\right].
\end{equation}
Finally, the horizontal components of the equation of motion at
$O(\epsilon^2)$ are
\begin{eqnarray}
  \lefteqn{\left(\f{\p}{\p\tau}-\rmi\Omega_2\right)v_{r0}'-\rmi\Omega_0v_{r2}'-2\Omega_2v_{\phi0}'-2\Omega_0v_{\phi2}'}&\nonumber\\
  &&=\f{3\Omega_0^2H_0H_0'}{r}-\f{1}{\Sigma_0}\f{\p P_0'}{\p r}+\f{\Sigma_0'}{\Sigma_0^2}\f{\rmd P_0}{\rmd r},
\end{eqnarray}
\begin{eqnarray}
  \lefteqn{\left(\f{\p}{\p\tau}-\rmi\Omega_2\right)v_{\phi0}'-\rmi\Omega_0v_{\phi2}'+\f{1}{2}\Omega_0v_{r2}'+\f{v_{r0}'}{r}\f{\rmd}{\rmd r}(r^2\Omega_2)}&\nonumber\\
  &&=\f{\rmi P_0'}{\Sigma_0r}.
\end{eqnarray}
We eliminate $v_{r2}'$ and $v_{\phi2}'$ by taking the $(1,2\rmi)$
linear combination of these equations:
\begin{eqnarray}
  \lefteqn{2\rmi r\Omega_0\f{\p E}{\p\tau}=2r^{1/2}\Omega_0E\f{\rmd}{\rmd r}(r^{3/2}\Omega_2)+\f{3\Omega_0^2H_0H_0'}{r}}&\nonumber\\
  &&-\f{1}{\Sigma_0r^2}\f{\p}{\p r}(P_0'r^2)+\f{\Sigma_0'}{\Sigma_0^2}\f{\rmd P_0}{\rmd r}.
\end{eqnarray}
Substituting for $\Omega_2$, $\Sigma_0'$, $P_0'$, $H_0'$ and
multiplying by $-\Sigma_0r$, we obtain
\begin{eqnarray}
  \lefteqn{-2\rmi\Sigma_0r^2\Omega_0\f{\p E}{\p\tau}=\Sigma_0E\f{\rmd}{\rmd r}\left(\f{3rP_0}{2\Sigma_0}-\f{r^2}{\Sigma_0}\f{\rmd P_0}{\rmd r}\right)}&\nonumber\\
  &&-3\Sigma_0\Omega_0^2ErH_0\f{\rmd H_0}{\rmd r}+\f{3}{\gamma}\Sigma_0\Omega_0^2H_0^2\left[3E-(\gamma-1)r\f{\p E}{\p r}\right]\nonumber\\
  &&+\f{1}{r}\f{\p}{\p r}\left\{Er^3\f{\rmd P_0}{\rmd r}+\f{P_0r^2}{\gamma}\left[3(\gamma-1)E+(2\gamma-1)r\f{\p E}{\p r}\right]\right\}\nonumber\\
  &&-\f{r^2}{\Sigma_0}\f{\rmd P_0}{\rmd r}\f{\p(\Sigma_0E)}{\p r},
\end{eqnarray}
which simplifies to
\begin{eqnarray}
  \lefteqn{-2\rmi\Sigma_0r^2\Omega_0\f{\p E}{\p\tau}=\f{1}{r}\f{\p}{\p r}\left[\left(2-\f{1}{\gamma}\right)P_0r^3\f{\p E}{\p r}\right]}&\nonumber\\
  &&+\left(4-\f{3}{\gamma}\right)r\f{\rmd P_0}{\rmd r}E+3\left(1+\f{1}{\gamma}\right)P_0E.
\end{eqnarray}
This Schr\"odinger-like dispersive wave equation agrees exactly with
the linear equation for the secular evolution of eccentricity in a 3D
adiabatic disc, as found in equation~(2) of
\citet{2016MNRAS.458.3221T} or equation~(176) of
\citet{2014MNRAS.445.2621O}.

\section{Conclusions}
\label{s:conclusions}

In this paper we have presented an affine model of the dynamics of
astrophysical discs. It extends the 2D hydrodynamic equations that are
often applied without adequate justification to thin discs. The
additional degrees of freedom included here allow the disc to expand
and contract in the vertical direction, to undergo deformation of the
midplane and to develop the internal shearing motions that accompany
such deformations. All of these are necessary to describe eccentric
and warped discs and we have shown that the model exactly reproduces
the linear secular theory of such discs in an appropriate
limit. However, it does not rely on any secular or small-amplitude
approximation and so should be useful in describing discs with general
combinations of tidal deformations, density waves, eccentricity and
warping. The equations of the affine model are 2D partial differential
equations that can be seen as a useful and generally more applicable
extension of the 2D hydrodynamic model.

The affine model is derived here, in the case of an ideal fluid, from
Hamilton's Principle after making a specific approximation to the
deformation gradient tensor. It naturally incorporates conservation
laws for total energy and potential vorticity (PV), even for
non-planar discs. We have therefore shown that PV or vortensity can be
defined for thin discs with variable thickness and with deformable
midplanes.

Future work should consider the numerical implementation of the
equations of the affine model and their application to various
astrophysical problems of interest. It would be valuable to include
non-ideal effects such as viscous or other shear stresses, heating and
cooling. It may also be possible to incorporate self-gravity and
magnetic fields in some approximation.

\section*{Acknowledgements}

This research was supported by STFC through grants ST/L000636/1 and
ST/P000673/1.  I am grateful to the referee for raising questions
about the dispersion relation that led to the modifications proposed
in Appendix~\ref{s:appendix2}.

\appendix
\onecolumn

\section{Equations of the affine model in Cartesian coordinates}
\label{s:appendix1}

In Cartesian coordinates the equations of Section~\ref{s:projected} read
\begin{eqnarray}
  \lefteqn{\left(\f{\p}{\p t}+v_x\f{\p}{\p X}+v_y\f{\p}{\p Y}\right)v_x=-\Phi_x-\left(\f{1}{2}H_x^2\Phi_{xxx}+\f{1}{2}H_y^2\Phi_{xyy}+\f{1}{2}H_z^2\Phi_{xzz}+H_xH_y\Phi_{xxy}+H_xH_z\Phi_{xxz}+H_yH_z\Phi_{xyz}\right)}&\nonumber\\
  &&-\f{1}{\Sigma}\f{\p P}{\p X}-\f{1}{\Sigma}\f{\p}{\p X}\left(\f{PH_x}{H}\f{\p Z}{\p X}\right)-\f{1}{\Sigma}\f{\p}{\p Y}\left(\f{PH_y}{H}\f{\p Z}{\p X}\right),
\end{eqnarray}
\begin{eqnarray}
  \lefteqn{\left(\f{\p}{\p t}+v_x\f{\p}{\p X}+v_y\f{\p}{\p Y}\right)v_y=-\Phi_y-\left(\f{1}{2}H_x^2\Phi_{xxy}+\f{1}{2}H_y^2\Phi_{yyy}+\f{1}{2}H_z^2\Phi_{yzz}+H_xH_y\Phi_{xyy}+H_xH_z\Phi_{xyz}+H_yH_z\Phi_{yyz}\right)}&\nonumber\\
  &&-\f{1}{\Sigma}\f{\p P}{\p Y}-\f{1}{\Sigma}\f{\p}{\p X}\left(\f{PH_x}{H}\f{\p Z}{\p Y}\right)-\f{1}{\Sigma}\f{\p}{\p Y}\left(\f{PH_y}{H}\f{\p Z}{\p Y}\right),
\end{eqnarray}
\begin{eqnarray}
  \lefteqn{\left(\f{\p}{\p t}+v_x\f{\p}{\p X}+v_y\f{\p}{\p Y}\right)v_z=-\Phi_z-\left(\f{1}{2}H_x^2\Phi_{xxz}+\f{1}{2}H_y^2\Phi_{yyz}+\f{1}{2}H_z^2\Phi_{zzz}+H_xH_y\Phi_{xyz}+H_xH_z\Phi_{xzz}+H_yH_z\Phi_{yzz}\right)}&\nonumber\\
  &&+\f{1}{\Sigma}\f{\p}{\p X}\left(\f{PH_x}{H}\right)+\f{1}{\Sigma}\f{\p}{\p Y}\left(\f{PH_y}{H}\right),
\end{eqnarray}
\begin{equation}
  \left(\f{\p}{\p t}+v_x\f{\p}{\p X}+v_y\f{\p}{\p Y}\right)w_x=-H_x\Phi_{xx}-H_y\Phi_{xy}-H_z\Phi_{xz}-\f{P}{\Sigma H}\f{\p Z}{\p X},
\end{equation}
\begin{equation}
  \left(\f{\p}{\p t}+v_x\f{\p}{\p X}+v_y\f{\p}{\p Y}\right)w_y=-H_x\Phi_{xy}-H_y\Phi_{yy}-H_z\Phi_{yz}-\f{P}{\Sigma H}\f{\p Z}{\p Y},
\end{equation}
\begin{equation}
  \left(\f{\p}{\p t}+v_x\f{\p}{\p X}+v_y\f{\p}{\p Y}\right)w_z=-H_x\Phi_{xz}-H_y\Phi_{xz}-H_z\Phi_{zz}+\f{P}{\Sigma H},
\end{equation}
with
\begin{equation}
  H=H_z-H_x\f{\p Z}{\p X}-H_y\f{\p Z}{\p Y},
\end{equation}
\begin{equation}
  v_z=\left(\f{\p}{\p t}+v_x\f{\p}{\p X}+v_y\f{\p}{\p Y}\right)Z,
\end{equation}
\begin{equation}
  w_x=\left(\f{\p}{\p t}+v_x\f{\p}{\p X}+v_y\f{\p}{\p Y}\right)H_x,
\end{equation}
\begin{equation}
  w_y=\left(\f{\p}{\p t}+v_x\f{\p}{\p X}+v_y\f{\p}{\p Y}\right)H_y,
\end{equation}
\begin{equation}
  w_z=\left(\f{\p}{\p t}+v_x\f{\p}{\p X}+v_y\f{\p}{\p Y}\right)H_z.
\end{equation}
Here
\begin{equation}
  \Phi_{xyz}=\f{\p^3\Phi}{\p x\,\p y\,\p z}\bigg|_{x=X,\,y=Y,\,z=Z}
\end{equation}
etc. The remaining equations can be written as, e.g.
\begin{equation}
  \left(\f{\p}{\p t}+v_x\f{\p}{\p X}+v_y\f{\p}{\p Y}\right)\Sigma=-\Sigma\left(\f{\p v_x}{\p X}+\f{\p v_y}{\p Y}\right),
\end{equation}
\begin{equation}
  \left(\f{\p}{\p t}+v_x\f{\p}{\p X}+v_y\f{\p}{\p Y}\right)P=-\gamma P\left(\f{\p v_x}{\p X}+\f{\p v_y}{\p Y}\right)-\f{(\gamma-1)P}{H}\f{\rmD H}{\rmD t},
\end{equation}
where
\begin{equation}
  \f{\rmD H}{\rmD t}=w_z-H_x\f{\p v_z}{\p X}-H_y\f{\p v_z}{\p Y}-\f{\p Z}{\p X}\left(w_x-H_x\f{\p v_x}{\p X}-H_y\f{\p v_x}{\p Y}\right)-\f{\p Z}{\p Y}\left(w_y-H_x\f{\p v_y}{\p X}-H_y\f{\p v_y}{\p Y}\right).
\end{equation}

In the symmetric case discussed in Section~\ref{s:symmetric}, these
equations reduce to
\begin{equation}
  \left(\f{\p}{\p t}+v_x\f{\p}{\p X}+v_y\f{\p}{\p Y}\right)v_x=-\Phi_x-\f{1}{2}H^2\Psi_x-\f{1}{\Sigma}\f{\p P}{\p X},
\end{equation}
\begin{equation}
  \left(\f{\p}{\p t}+v_x\f{\p}{\p X}+v_y\f{\p}{\p Y}\right)v_y=-\Phi_y-\f{1}{2}H^2\Psi_y-\f{1}{\Sigma}\f{\p P}{\p Y},
\end{equation}
\begin{equation}
 \left(\f{\p}{\p t}+v_x\f{\p}{\p X}+v_y\f{\p}{\p Y}\right)w=-\Psi H+\f{P}{\Sigma H},
\end{equation}
\begin{equation}
  \left(\f{\p}{\p t}+v_x\f{\p}{\p X}+v_y\f{\p}{\p Y}\right)H=w,
\end{equation}
\begin{equation}
  \left(\f{\p}{\p t}+v_x\f{\p}{\p X}+v_y\f{\p}{\p Y}\right)\Sigma=-\Sigma\left(\f{\p v_x}{\p X}+\f{\p v_y}{\p Y}\right),
\end{equation}
\begin{equation}
  \left(\f{\p}{\p t}+v_x\f{\p}{\p X}+v_y\f{\p}{\p Y}\right)P=-\gamma P\left(\f{\p v_x}{\p X}+\f{\p v_y}{\p Y}\right)-\f{(\gamma-1)Pw}{H},
\end{equation}
where
\begin{equation}
  \Psi=\f{\p^2\Phi}{\p z^2}\bigg|_{z=0}.
\end{equation}

\section{Additional terms resulting from extensions of the thin-disc
  approximation}
\label{s:appendix2}

The exact Jacobian determinant of the second stage of the map
(equation~\ref{j_second}) is a quadratic function of $\zeta$ in which
the terms dependent on $\zeta$ involve spatial derivatives of $\bmH$.
The approximate expression $H$ is subject to the correction factor
\begin{equation}
  1+\left\{\bar\grad\bcdot\bar\bmH-\f{1}{H}\bmn\bcdot[(\bar\bmH\bcdot\bar\grad)\bmH]\right\}\zeta+\f{\bmH}{H}\bcdot\left(\f{\p\bmH}{\p X}\btimes\f{\p\bmH}{\p Y}\right)\zeta^2.
\label{correction}
\end{equation}
These terms become important when the condition $\|\bar\grad\bmH\|$ is
not satisfied, and the assumption of a uniform expansion or
contraction of the fluid columns is violated.  In principle, the exact
internal energy associated with the affine transformation could be
computed, as a function of $\bar\grad\bmH$, by raising this expression
to the power $-(\gamma-1)$ and integrating over $\zeta$, weighted by
$F_p(\zeta)$.  We will not pursue this approach because the complexity
it introduces is not justified by the simplicity of our assumption
regarding the affine transformation.

Let us write the correction factor (after averaging over $\zeta$) as
$\rme^F$, where $F$ depends on the spatial derivatives of $\bmH$; it
might also depend on $\bmH$ itself and (through $\bmn$) on the spatial
derivatives of $Z$.  We then have
\begin{equation}
  J_3=J_2\f{H}{H_0}\rme^F,
\end{equation}
and the new factor of $\rme^{-(\gamma-1)F}$ in the internal energy
contribution to the Lagrangian gives rise to the following additional
terms in the equations of motion:
\begin{equation}
   \f{\rmD^2\bar X_i}{\rmD t^2}=\cdots+\f{1}{\Sigma}\f{\p}{\p\bar X_j}\left[P\f{\p H_k}{\p\bar X_i}\f{\p F}{\p(\p H_k/\p\bar X_j)}\right]+\f{1}{\Sigma}\f{\p}{\p\bar X_j}\left[P\f{\p Z}{\p\bar X_i}\f{\p F}{\p(\p Z/\p\bar X_j)}\right],
\end{equation}
\begin{equation}
   \f{\rmD^2Z}{\rmD t^2}=\cdots-\f{1}{\Sigma}\f{\p}{\p\bar X_j}\left[P\f{\p F}{\p(\p Z/\p\bar X_j)}\right],
\end{equation}
\begin{equation}
   \f{\rmD^2H_i}{\rmD t^2}=\cdots+\f{P}{\Sigma}\f{\p F}{\p H_i}-\f{1}{\Sigma}\f{\p}{\p\bar X_j}\left[P\f{\p F}{\p(\p H_i/\p\bar X_j)}\right],
\end{equation}
where summation over $j=\{1,2\}$ is implied.  The evolutionary
equation for $P$ is also modified to
\begin{equation}
  \f{\rmD P}{\rmD t}=\cdots-(\gamma-1)P\f{\rmD F}{\rmD t},
\end{equation}
which, together with the modified equations for $\bar X_i$ and $\bar
H_i$, conserves the total energy in the same form as
equation~(\ref{energy}).\footnote{Note that the `novel' terms
  involving $\bmn$, already present in equations (\ref{d2r}) and
  (\ref{d2a}), derive from the above rules applied to the function
  $F=\ln H=\ln(\bmH\bcdot\bmn)$, which describes the departure of
  $J_3$ from $J_2$ in our standard affine model.}

A useful model is
\begin{equation}
  F_1=-\f{1}{2}(\bar\grad\bcdot\bar\bmH)^2.
\label{model1}
\end{equation}
This is motivated by the first correction term
$(\bar\grad\bcdot\bar\bmH)\zeta$ in equation~(\ref{correction}), which
is relatively easy to understand.  If $H_x$ increases with $X$, for
example, then the variable tilt of the fluid columns rarifies the disc
above the midplane and compresses it below.  The increase in net
internal energy at second order is here modelled by the factor
$\rme^{-(\gamma-1)F_1}$.  This model produces the following additional
terms:
\begin{equation}
  \f{\rmD^2\bar X_i}{\rmD t^2}=\cdots-\f{1}{\Sigma}\f{\p}{\p\bar X_j}\left[P(\bar\grad\bcdot\bar\bmH)\f{\p\bar H_j}{\p\bar X_i}\right],
\end{equation}
\begin{equation}
  \f{\rmD^2\bar\bmH}{\rmD t^2}=\cdots+\f{1}{\Sigma}\bar\grad[P(\bar\grad\bcdot\bar\bmH)],
\end{equation}
\begin{equation}
  \f{\rmD P}{\rmD t}=\cdots+(\gamma-1)P(\bar\grad\bcdot\bar\bmH)\left(\bar\grad\bcdot\bar\bmw-\f{\p\bar v_i}{\p\bar X_j}\f{\p\bar H_j}{\p\bar X_i}\right).
\end{equation}

A second useful modification is to add
\begin{equation}
  F_2=-\f{1}{2}|\bar\grad H_z|^2.
\label{model2}
\end{equation}
Although harder to justify based on equation~(\ref{correction}), this
model produces the following additional terms that are found to
improve the dispersion relation of symmetric modes at short
wavelengths:
\begin{equation}
  \f{\rmD^2\bar X_i}{\rmD t^2}=\cdots-\f{1}{\Sigma}\f{\p}{\p\bar X_j}\left(P\f{\p H_z}{\p\bar X_j}\f{\p H_z}{\p\bar X_i}\right),
\end{equation}
\begin{equation}
  \f{\rmD^2H_z}{\rmD t^2}=\cdots+\f{1}{\Sigma}\bar\grad\bcdot(P\bar\grad H_z),
\end{equation}
\begin{equation}
  \f{\rmD P}{\rmD t}=\cdots+(\gamma-1)P(\bar\grad H_z)\bcdot\left(\bar\grad w_z-\f{\p H_z}{\p\bar X_i}\bar\grad\bar v_i\right).
\end{equation}

\label{lastpage}

\end{document}